\documentclass[letterpaper,journal]{IEEEtran}

\usepackage{amsmath,graphicx}

\usepackage[T1]{fontenc}
\usepackage{graphicx}
\usepackage{amssymb}
\usepackage{amsmath}
\usepackage{amsthm}
\usepackage{subfigure}
\usepackage{booktabs} 
\usepackage{multirow}
\usepackage{microtype}
\usepackage{cite}

\usepackage{url}
\usepackage{balance}
\usepackage{colortbl}

\usepackage{tikz}

\usepackage{amssymb}
\usepackage{amsfonts}
\usepackage{mathrsfs}
\usepackage{xspace}
\usepackage{bm}
\usepackage{upgreek}

\newcommand{\safemath}[2]{\newcommand{#1}{\ensuremath{#2}\xspace}}

\safemath{\bma}{\mathbf{a}}
\safemath{\bmb}{\mathbf{b}}
\safemath{\bmc}{\mathbf{c}}
\safemath{\bmd}{\mathbf{d}}
\safemath{\bme}{\mathbf{e}}
\safemath{\bmf}{\mathbf{f}}
\safemath{\bmg}{\mathbf{g}}
\safemath{\bmh}{\mathbf{h}}
\safemath{\bmi}{\mathbf{i}}
\safemath{\bmj}{\mathbf{j}}
\safemath{\bmk}{\mathbf{k}}
\safemath{\bml}{\mathbf{l}}
\safemath{\bmm}{\mathbf{m}}
\safemath{\bmn}{\mathbf{n}}
\safemath{\bmo}{\mathbf{o}}
\safemath{\bmp}{\mathbf{p}}
\safemath{\bmq}{\mathbf{q}}
\safemath{\bmr}{\mathbf{r}}
\safemath{\bms}{\mathbf{s}}
\safemath{\bmt}{\mathbf{t}}
\safemath{\bmu}{\mathbf{u}}
\safemath{\bmv}{\mathbf{v}}
\safemath{\bmw}{\mathbf{w}}
\safemath{\bmx}{\mathbf{x}}
\safemath{\bmy}{\mathbf{y}}
\safemath{\bmz}{\mathbf{z}}
\safemath{\bmzero}{\mathbf{0}}
\safemath{\bmone}{\mathbf{1}}
\safemath{\Bell}{\ensuremath{\boldsymbol\ell}}

\bmdefine{\biad}{a}
\bmdefine{\bibd}{b}
\bmdefine{\bicd}{c}
\bmdefine{\bidd}{d}
\bmdefine{\bied}{e}
\bmdefine{\bifd}{f}
\bmdefine{\bigd}{g}
\bmdefine{\bihd}{h}
\bmdefine{\biid}{i}
\bmdefine{\bijd}{j}
\bmdefine{\bikd}{k}
\bmdefine{\bild}{l}
\bmdefine{\bimd}{m}
\bmdefine{\bind}{n}
\bmdefine{\biod}{o}
\bmdefine{\bipd}{p}
\bmdefine{\biqd}{q}
\bmdefine{\bird}{r}
\bmdefine{\bisd}{s}
\bmdefine{\bitd}{t}
\bmdefine{\biud}{u}
\bmdefine{\bivd}{v}
\bmdefine{\biwd}{w}
\bmdefine{\bixd}{x}
\bmdefine{\biyd}{y}
\bmdefine{\bizd}{z}

\bmdefine{\bixid}{\xi}
\bmdefine{\bilambdad}{\lambda}
\bmdefine{\bimud}{\mu}
\bmdefine{\bithetad}{\theta}
\bmdefine{\biphid}{\phi}
\bmdefine{\bideltad}{\delta}

\safemath{\bmia}{\biad}
\safemath{\bmib}{\bibd}
\safemath{\bmic}{\bicd}
\safemath{\bmid}{\bidd}
\safemath{\bmie}{\bied}
\safemath{\bmif}{\bifd}
\safemath{\bmig}{\bigd}
\safemath{\bmih}{\bihd}
\safemath{\bmii}{\biid}
\safemath{\bmij}{\bijd}
\safemath{\bmik}{\bikd}
\safemath{\bmil}{\bild}
\safemath{\bmim}{\bimd}
\safemath{\bmin}{\bind}
\safemath{\bmio}{\biod}
\safemath{\bmip}{\bipd}
\safemath{\bmiq}{\biqd}
\safemath{\bmir}{\bird}
\safemath{\bmis}{\bisd}
\safemath{\bmit}{\bitd}
\safemath{\bmiu}{\biud}
\safemath{\bmiv}{\bivd}
\safemath{\bmiw}{\biwd}
\safemath{\bmix}{\bixd}
\safemath{\bmiy}{\biyd}
\safemath{\bmiz}{\bizd}

\safemath{\bmxi}{\bixid}
\safemath{\bmlambda}{\bilambdad}
\safemath{\bmmu}{\bimud}
\safemath{\bmtheta}{\bithetad}
\safemath{\bmphi}{\biphid}
\safemath{\bmdelta}{\bideltad}

\safemath{\bA}{\mathbf{A}}
\safemath{\bB}{\mathbf{B}}
\safemath{\bC}{\mathbf{C}}
\safemath{\bD}{\mathbf{D}}
\safemath{\bE}{\mathbf{E}}
\safemath{\bF}{\mathbf{F}}
\safemath{\bG}{\mathbf{G}}
\safemath{\bH}{\mathbf{H}}
\safemath{\bI}{\mathbf{I}}
\safemath{\bJ}{\mathbf{J}}
\safemath{\bK}{\mathbf{K}}
\safemath{\bL}{\mathbf{L}}
\safemath{\bM}{\mathbf{M}}
\safemath{\bN}{\mathbf{N}}
\safemath{\bO}{\mathbf{O}}
\safemath{\bP}{\mathbf{P}}
\safemath{\bQ}{\mathbf{Q}}
\safemath{\bR}{\mathbf{R}}
\safemath{\bS}{\mathbf{S}}
\safemath{\bT}{\mathbf{T}}
\safemath{\bU}{\mathbf{U}}
\safemath{\bV}{\mathbf{V}}
\safemath{\bW}{\mathbf{W}}
\safemath{\bX}{\mathbf{X}}
\safemath{\bY}{\mathbf{Y}}
\safemath{\bZ}{\mathbf{Z}}

\safemath{\bZero}{\mathbf{0}}
\safemath{\bOne}{\mathbf{1}}
\safemath{\bDelta}{\mathbf{\Delta}}
\safemath{\bLambda}{\mathbf{\UpLambda}}
\safemath{\bPhi}{\mathbf{\Upphi}}
\safemath{\bSigma}{\mathbf{\Upsigma}}
\safemath{\bOmega}{\mathbf{\Upomega}}
\safemath{\bTheta}{\mathbf{\Uptheta}}

\bmdefine{\biAd}{A}
\bmdefine{\biBd}{B}
\bmdefine{\biCd}{C}
\bmdefine{\biDd}{D}
\bmdefine{\biEd}{E}
\bmdefine{\biFd}{F}
\bmdefine{\biGd}{G}
\bmdefine{\biHd}{H}
\bmdefine{\biId}{I}
\bmdefine{\biJd}{J}
\bmdefine{\biKd}{K}
\bmdefine{\biLd}{L}
\bmdefine{\biMd}{M}
\bmdefine{\biOd}{N}
\bmdefine{\biPd}{O}
\bmdefine{\biQd}{P}
\bmdefine{\biRd}{R}
\bmdefine{\biSd}{S}
\bmdefine{\biTd}{T}
\bmdefine{\biUd}{U}
\bmdefine{\biVd}{V}
\bmdefine{\biWd}{W}
\bmdefine{\biXd}{X}
\bmdefine{\biYd}{Y}
\bmdefine{\biZd}{Z}

\bmdefine{\biDelta}{\Delta}
\bmdefine{\biLambda}{\Lambda}
\bmdefine{\biPhi}{\Phi}
\bmdefine{\biSigma}{\Sigma}
\bmdefine{\biOmega}{\Omega}
\bmdefine{\biTheta}{\Theta}

\safemath{\bimA}{\biAd}
\safemath{\bimB}{\biBd}
\safemath{\bimC}{\biCd}
\safemath{\bimD}{\biDd}
\safemath{\bimE}{\biEd}
\safemath{\bimF}{\biFd}
\safemath{\bimG}{\biGd}
\safemath{\bimH}{\biHd}
\safemath{\bimI}{\biId}
\safemath{\bimJ}{\biJd}
\safemath{\bimK}{\biKd}
\safemath{\bimL}{\biLd}
\safemath{\bimM}{\biMd}
\safemath{\bimN}{\biNd}
\safemath{\bimO}{\biOd}
\safemath{\bimP}{\biPd}
\safemath{\bimQ}{\biQd}
\safemath{\bimR}{\biRd}
\safemath{\bimS}{\biSd}
\safemath{\bimT}{\biTd}
\safemath{\bimU}{\biUd}
\safemath{\bimV}{\biVd}
\safemath{\bimW}{\biWd}
\safemath{\bimX}{\biXd}
\safemath{\bimY}{\biYd}
\safemath{\bimZ}{\biZd}

\safemath{\bimDelta}{\biDelta}
\safemath{\bimLambda}{\biLambda}
\safemath{\bimPhi}{\biPhi}
\safemath{\bimSigma}{\biSigma}
\safemath{\bimOmega}{\biOmega}
\safemath{\bimTheta}{\biTheta}

\safemath{\setA}{\mathcal{A}}
\safemath{\setB}{\mathcal{B}}
\safemath{\setC}{\mathcal{C}}
\safemath{\setD}{\mathcal{D}}
\safemath{\setE}{\mathcal{E}}
\safemath{\setF}{\mathcal{F}}
\safemath{\setG}{\mathcal{G}}
\safemath{\setH}{\mathcal{H}}
\safemath{\setI}{\mathcal{I}}
\safemath{\setJ}{\mathcal{J}}
\safemath{\setK}{\mathcal{K}}
\safemath{\setL}{\mathcal{L}}
\safemath{\setM}{\mathcal{M}}
\safemath{\setN}{\mathcal{N}}
\safemath{\setO}{\mathcal{O}}
\safemath{\setP}{\mathcal{P}}
\safemath{\setQ}{\mathcal{Q}}
\safemath{\setR}{\mathcal{R}}
\safemath{\setS}{\mathcal{S}}
\safemath{\setT}{\mathcal{T}}
\safemath{\setU}{\mathcal{U}}
\safemath{\setV}{\mathcal{V}}
\safemath{\setW}{\mathcal{W}}
\safemath{\setX}{\mathcal{X}}
\safemath{\setY}{\mathcal{Y}}
\safemath{\setZ}{\mathcal{Z}}
\safemath{\emptySet}{\varnothing}

\safemath{\colA}{\mathscr{A}}
\safemath{\colB}{\mathscr{B}}
\safemath{\colC}{\mathscr{C}}
\safemath{\colD}{\mathscr{D}}
\safemath{\colE}{\mathscr{E}}
\safemath{\colF}{\mathscr{F}}
\safemath{\colG}{\mathscr{G}}
\safemath{\colH}{\mathscr{H}}
\safemath{\colI}{\mathscr{I}}
\safemath{\colJ}{\mathscr{J}}
\safemath{\colK}{\mathscr{K}}
\safemath{\colL}{\mathscr{L}}
\safemath{\colM}{\mathscr{M}}
\safemath{\colN}{\mathscr{N}}
\safemath{\colO}{\mathscr{O}}
\safemath{\colP}{\mathscr{P}}
\safemath{\colQ}{\mathscr{Q}}
\safemath{\colR}{\mathscr{R}}
\safemath{\colS}{\mathscr{S}}
\safemath{\colT}{\mathscr{T}}
\safemath{\colU}{\mathscr{U}}
\safemath{\colV}{\mathscr{V}}
\safemath{\colW}{\mathscr{W}}
\safemath{\colX}{\mathscr{X}}
\safemath{\colY}{\mathscr{Y}}
\safemath{\colZ}{\mathscr{Z}}

\safemath{\opA}{\mathbb{A}}
\safemath{\opB}{\mathbb{B}}
\safemath{\opC}{\mathbb{C}}
\safemath{\opD}{\mathbb{D}}
\safemath{\opE}{\mathbb{E}}
\safemath{\opF}{\mathbb{F}}
\safemath{\opG}{\mathbb{G}}
\safemath{\opH}{\mathbb{H}}
\safemath{\opI}{\mathbb{I}}
\safemath{\opJ}{\mathbb{J}}
\safemath{\opK}{\mathbb{K}}
\safemath{\opL}{\mathbb{L}}
\safemath{\opM}{\mathbb{M}}
\safemath{\opN}{\mathbb{N}}
\safemath{\opO}{\mathbb{O}}
\safemath{\opP}{\mathbb{P}}
\safemath{\opQ}{\mathbb{Q}}
\safemath{\opR}{\mathbb{R}}
\safemath{\opS}{\mathbb{S}}
\safemath{\opT}{\mathbb{T}}
\safemath{\opU}{\mathbb{U}}
\safemath{\opV}{\mathbb{V}}
\safemath{\opW}{\mathbb{W}}
\safemath{\opX}{\mathbb{X}}
\safemath{\opY}{\mathbb{Y}}
\safemath{\opZ}{\mathbb{Z}}
\safemath{\opZero}{\mathbb{O}}
\safemath{\identityop}{\opI}

\safemath{\veca}{\bma}
\safemath{\vecb}{\bmb}
\safemath{\vecc}{\bmc}
\safemath{\vecd}{\bmd}
\safemath{\vece}{\bme}
\safemath{\vecf}{\bmf}
\safemath{\vecg}{\bmg}
\safemath{\vech}{\bmh}
\safemath{\veci}{\bmi}
\safemath{\vecj}{\bmj}
\safemath{\veck}{\bmk}
\safemath{\vecl}{\bml}
\safemath{\vecm}{\bmm}
\safemath{\vecn}{\bmn}
\safemath{\veco}{\bmo}
\safemath{\vecp}{\bmp}
\safemath{\vecq}{\bmq}
\safemath{\vecr}{\bmr}
\safemath{\vecs}{\bms}
\safemath{\vect}{\bmt}
\safemath{\vecu}{\bmu}
\safemath{\vecv}{\bmv}
\safemath{\vecw}{\bmw}
\safemath{\vecx}{\bmx}
\safemath{\vecy}{\bmy}
\safemath{\vecz}{\bmz}

\safemath{\veczero}{\bmzero}
\safemath{\vecone}{\bmone}
\safemath{\vecxi}{\bmxi}
\safemath{\veclambda}{\bmlambda}
\safemath{\vecmu}{\bmmu}
\safemath{\vectheta}{\bmtheta}
\safemath{\vecphi}{\bmphi}
\safemath{\vecdelta}{\bmdelta}

\safemath{\matA}{\bA}
\safemath{\matB}{\bB}
\safemath{\matC}{\bC}
\safemath{\matD}{\bD}
\safemath{\matE}{\bE}
\safemath{\matF}{\bF}
\safemath{\matG}{\bG}
\safemath{\matH}{\bH}
\safemath{\matI}{\bI}
\safemath{\matJ}{\bJ}
\safemath{\matK}{\bK}
\safemath{\matL}{\bL}
\safemath{\matM}{\bM}
\safemath{\matN}{\bN}
\safemath{\matO}{\bO}
\safemath{\matP}{\bP}
\safemath{\matQ}{\bQ}
\safemath{\matR}{\bR}
\safemath{\matS}{\bS}
\safemath{\matT}{\bT}
\safemath{\matU}{\bU}
\safemath{\matV}{\bV}
\safemath{\matW}{\bW}
\safemath{\matX}{\bX}
\safemath{\matY}{\bY}
\safemath{\matZ}{\bZ}
\safemath{\matzero}{\bmzero}

\safemath{\matDelta}{\bDelta}
\safemath{\matLambda}{\bLambda}
\safemath{\matPhi}{\bPhi}
\safemath{\matSigma}{\bSigma}
\safemath{\matOmega}{\bOmega}
\safemath{\matTheta}{\bTheta}

\safemath{\matidentity}{\matI}
\safemath{\matone}{\matO}

\safemath{\rnda}{A}
\safemath{\rndb}{B}
\safemath{\rndc}{C}
\safemath{\rndd}{D}
\safemath{\rnde}{E}
\safemath{\rndf}{F}
\safemath{\rndg}{G}
\safemath{\rndh}{H}
\safemath{\rndi}{I}
\safemath{\rndj}{J}
\safemath{\rndk}{K}
\safemath{\rndl}{L}
\safemath{\rndm}{M}
\safemath{\rndn}{N}
\safemath{\rndo}{O}
\safemath{\rndp}{P}
\safemath{\rndq}{Q}
\safemath{\rndr}{R}
\safemath{\rnds}{S}
\safemath{\rndt}{T}
\safemath{\rndu}{U}
\safemath{\rndv}{V}
\safemath{\rndw}{W}
\safemath{\rndx}{X}
\safemath{\rndy}{Y}
\safemath{\rndz}{Z}

\safemath{\rveca}{\bimA}
\safemath{\rvecb}{\bimB}
\safemath{\rvecc}{\bimC}
\safemath{\rvecd}{\bimD}
\safemath{\rvece}{\bimE}
\safemath{\rvecf}{\bimF}
\safemath{\rvecg}{\bimG}
\safemath{\rvech}{\bimH}
\safemath{\rveci}{\bimI}
\safemath{\rvecj}{\bimJ}
\safemath{\rveck}{\bimK}
\safemath{\rvecl}{\bimL}
\safemath{\rvecm}{\bimM}
\safemath{\rvecn}{\bimN}
\safemath{\rveco}{\bomO}
\safemath{\rvecp}{\bimP}
\safemath{\rvecq}{\bimQ}
\safemath{\rvecr}{\bimR}
\safemath{\rvecs}{\bimS}
\safemath{\rvect}{\bimT}
\safemath{\rvecu}{\bimU}
\safemath{\rvecv}{\bimV}
\safemath{\rvecw}{\bimW}
\safemath{\rvecx}{\bimX}
\safemath{\rvecy}{\bimY}
\safemath{\rvecz}{\bimZ}

\safemath{\rvecxi}{\bmxi}
\safemath{\rveclambda}{\bmlambda}
\safemath{\rvecmu}{\bmmu}
\safemath{\rvectheta}{\bmtheta}
\safemath{\rvecphi}{\bmphi}

\safemath{\rmatA}{\bimA}
\safemath{\rmatB}{\bimB}
\safemath{\rmatC}{\bimC}
\safemath{\rmatD}{\bimD}
\safemath{\rmatE}{\bimE}
\safemath{\rmatF}{\bimF}
\safemath{\rmatG}{\bimG}
\safemath{\rmatH}{\bimH}
\safemath{\rmatI}{\bimI}
\safemath{\rmatJ}{\bimJ}
\safemath{\rmatK}{\bimK}
\safemath{\rmatL}{\bimL}
\safemath{\rmatM}{\bimM}
\safemath{\rmatN}{\bimN}
\safemath{\rmatO}{\bimO}
\safemath{\rmatP}{\bimP}
\safemath{\rmatQ}{\bimQ}
\safemath{\rmatR}{\bimR}
\safemath{\rmatS}{\bimS}
\safemath{\rmatT}{\bimT}
\safemath{\rmatU}{\bimU}
\safemath{\rmatV}{\bimV}
\safemath{\rmatW}{\bimW}
\safemath{\rmatX}{\bimX}
\safemath{\rmatY}{\bimY}
\safemath{\rmatZ}{\bimZ}

\safemath{\rmatDelta}{\bimDelta}
\safemath{\rmatLambda}{\bimLambda}
\safemath{\rmatPhi}{\bimPhi}
\safemath{\rmatSigma}{\bimSigma}
\safemath{\rmatOmega}{\bimOmega}
\safemath{\rmatTheta}{\bimTheta}

\usepackage{amssymb}
\usepackage{amsfonts}
\usepackage{mathrsfs}
\usepackage{xspace}
\usepackage{bm}
\usepackage{fancyref}
\usepackage{textcomp}

\usepackage{multirow}
\usepackage{stmaryrd}

\newenvironment{textbmatrix}{	\setlength{\arraycolsep}{2.5pt}%
								\left[\begin{matrix}}{\end{matrix}\right]%
								\raisebox{0.08ex}{\vphantom{M}}}

\def\be{\begin{equation}}
\def\ee{\end{equation}}
\def\een{\nonumber \end{equation}}
\def\mat{\begin{bmatrix}}
\def\emat{\end{bmatrix}}
\def\btm{\begin{textbmatrix}}
\def\etm{\end{textbmatrix}}

\def\ba#1\ea{\begin{align}#1\end{align}}
\def\bas#1\eas{\begin{align*}#1\end{align*}}
\def\bs#1\es{\begin{split}#1\end{split}}
\def\bg#1\eg{\begin{gather}#1\end{gather}}
\def\bml#1\eml{\begin{multline}#1\end{multline}}
\def\bi#1\ei{\begin{itemize}#1\end{itemize}}

\newcommand{\lefto}{\mathopen{}\left}

\DeclareMathOperator*{\argmax}{arg\;max}		% arg max
\DeclareMathOperator{\Exop}{\opE}			% expectation operator
\newcommand{\Ex}[1]{\ensuremath{\Exop\lefto[#1\right]}} 	% expectation
\safemath{\dirac}{\delta}					% Dirac delta
\safemath{\krond}{\dirac}					% Kronecker delta
\safemath{\upto}{\uparrow}
\safemath{\downto}{\downarrow}
\safemath{\iu}{j}							% imaginary unit
\safemath{\ev}{\lambda}						% eigenvalue
\safemath{\hilseqspace}{l^{2}}				% Hilbert sequence space
\newcommand{\banachfunspace}[1]{\setL^{#1}}	% Banach function space
\safemath{\hilfunspace}{\banachfunspace{2}}	% Hilbert function space
\safemath{\SNR}{\textit{SNR}} 				% signal to noise ratio
\safemath{\PAR}{\textit{PAR}} 				% signal to noise ratio
\safemath{\No}{N_0}							% noise spectral density
\safemath{\Es}{E_s}							% energy per symbol
\safemath{\Eb}{E_b}							% energy per bit
\safemath{\EbNo}{\frac{\Eb}{\No}}
\safemath{\EsNo}{\frac{\Es}{\No}}

\DeclareMathOperator{\CHop}{\ensuremath{\opH}} % channel operator
\safemath{\tvir}{\rndh_{\CHop}}				% time-varying impulse response
\safemath{\tvtf}{\rndl_{\CHop}}				% 	-''- transfer function
\safemath{\spf}{\rnds_{\CHop}}				% spreading function
\safemath{\bff}{H_{\CHop}}					% bi-freuqency function

\safemath{\ircf}{r_{h}}						% impulse response correlation fn.
\safemath{\tftvcf}{r_{s}}					% scattering function
\safemath{\tfcf}{r_{l}}						% time-frequency correlation fn.
\safemath{\bfcf}{r_{H}}						% bi-frequency correlation fn.

\safemath{\tcorr}{c_h}						% time-correlation function
\safemath{\scf}{c_{s}}						% spreading function
\safemath{\tfcorr}{c_{l}}					% transfer-function correlation
\safemath{\fcorr}{c_{H}}						% frequency-correlation function

\safemath{\mi}{I}							% mutual information
\safemath{\capacity}{C}						% capacity

\safemath{\normal}{\mathcal{N}}			% normal distribution
\safemath{\jpg}{\mathcal{CN}}			% jointly proper Gaussian
\safemath{\mchain}{\leftrightarrow}		% Markov chain
\safemath{\dB}{\,\mathrm{dB}}
\safemath{\dBm}{\,\mathrm{dBm}}
\safemath{\Hz}{\,\mathrm{Hz}}
\safemath{\kHz}{\,\mathrm{kHz}}
\safemath{\MHz}{\,\mathrm{MHz}}
\safemath{\GHz}{\,\mathrm{GHz}}
\safemath{\s}{\,\mathrm{s}}
\safemath{\ms}{\,\mathrm{ms}}
\safemath{\mus}{\,\mathrm{\text{\textmu}s}}
\safemath{\ns}{\,\mathrm{ns}}
\safemath{\ps}{\,\mathrm{ps}}
\safemath{\meter}{\,\mathrm{m}}
\safemath{\mm}{\,\mathrm{mm}}
\safemath{\cm}{\,\mathrm{cm}}
\safemath{\m}{\,\mathrm{m}}
\safemath{\W}{\,\mathrm{W}}
\safemath{\mW}{\, \mathrm{mW}}
\safemath{\J}{\,\mathrm{J}}
\safemath{\K}{\,\mathrm{K}}
\safemath{\bit}{\,\mathrm{bit}}
\safemath{\nat}{\,\mathrm{nat}}

\safemath{\define}{\triangleq}			% definition

\safemath{\equivalent}{\sim}
\safemath{\distas}{\sim}					% distributed according to
\safemath{\sdiff}{\Delta}				% symmetric set difference

\safemath{\reals}{\mathbb{R}}
\safemath{\positivereals}{\reals_{+}}
\safemath{\integers}{\mathbb{Z}}
\safemath{\posint}{\integers_{+}}
\safemath{\naturals}{\mathbb{N}}
\safemath{\posnaturals}{\naturals_{+}}
\safemath{\complexset}{\mathbb{C}}
\safemath{\rationals}{\mathbb{Q}}

\newcommand*{\fancyrefapplabelprefix}{app}		% Appendix
\newcommand*{\fancyrefthmlabelprefix}{thm}		% Theorem
\newcommand*{\fancyreflemlabelprefix}{lem}		% Lemma
\newcommand*{\fancyrefcorlabelprefix}{cor}		% Corollary
\newcommand*{\fancyrefdeflabelprefix}{def}		% Definition
\newcommand*{\fancyrefproplabelprefix}{prop}		% Proposition
\newcommand*{\fancyrefexmpllabelprefix}{exmpl}
\newcommand*{\fancyrefalglabelprefix}{alg}		% Algorithm
\newcommand*{\fancyreftbllabelprefix}{tbl}		% Algorithm

\frefformat{vario}{\fancyrefseclabelprefix}{Sec.~#1}
\frefformat{vario}{\fancyrefthmlabelprefix}{Theorem~#1}
\frefformat{vario}{\fancyreftbllabelprefix}{Tbl.~#1}
\frefformat{vario}{\fancyreflemlabelprefix}{Lemma~#1}
\frefformat{vario}{\fancyrefcorlabelprefix}{Corollary~#1}
\frefformat{vario}{\fancyrefdeflabelprefix}{Definition~#1}
\frefformat{vario}{\fancyreffiglabelprefix}{Fig.~#1}
\frefformat{vario}{\fancyrefapplabelprefix}{Appendix~#1}
\frefformat{vario}{\fancyrefeqlabelprefix}{(#1)}
\frefformat{vario}{\fancyrefproplabelprefix}{Proposition~#1}
\frefformat{vario}{\fancyrefexmpllabelprefix}{Example~#1}
\frefformat{vario}{\fancyrefalglabelprefix}{Algorithm~#1}

\safemath{\dictab}{[\,\dicta\,\,\dictb\,]}

\safemath{\ysig}{\bmy}
\safemath{\ysighat}{\hat{\ysig}}
\safemath{\ysigdim}{M}
\safemath{\xsig}{\bmx}
\safemath{\xsigdim}{N}
\safemath{\nx}{n_x}
\safemath{\zsig}{\bmz}
\safemath{\zsigdim}{\ysigdim}
\safemath{\rsig}{\bmr}
\safemath{\Adict}{\bA}
\safemath{\Adicttilde}{\widetilde{\Adict}}
\safemath{\Adictdim}{\outputdim\times\xsigdim}
\safemath{\avec}{\bma}
\safemath{\avectilde}{\tilde{\avec}}
\safemath{\Bdict}{\bB}
\safemath{\Bdicttilde}{\widetilde{\Bdict}}
\safemath{\Cdict}{\bC}
\safemath{\cvec}{\bmc}
\safemath{\Ddict}{\bD}
\safemath{\Ddictdim}{\ysigdim\times\xsigdim}
\safemath{\dvec}{\bmd}
\safemath{\Ddicttilde}{\widetilde{\bD}}
\safemath{\Bonb}{\bB}
\safemath{\bvec}{\bmb}
\safemath{\Bonbdim}{\ysigdim\times\ysigdim}
\safemath{\noise}{\bmn}
\safemath{\noisedim}{\ysigim}
\safemath{\err}{\bme}
\safemath{\errdim}{\ysigdim}
\safemath{\errset}{\setE}
\safemath{\nerr}{n_e}
\safemath{\delop}{\bP_\errset}
\safemath{\delopc}{\bP_{{\errset}^c}}

\safemath{\cplxi}{\imath}
\safemath{\cplxj}{\jmath}
\safemath{\dict}{\matD}
\safemath{\inputdim}{N}		% number of columns of dictionary D
\safemath{\outputdim}{M}		%number of rows of dictionary D
\safemath{\sparsity}{S}	%sparsity
\safemath{\inputdimA}{{N_a}}	%total number of elements in dictionary A
\safemath{\inputdimB}{{N_b}}	%total number of elements in dictionary B
\safemath{\elemA}{{n_a}}	%number of elements chosen from dictionary A
\safemath{\elemB}{{n_b}}	%number of elements chosen from dictionary B
\safemath{\resA}{\matR_a}	%restriction map to elements of dictionary A
\safemath{\resB}{\matR_b}	%restriction map to elements of dictionary B
\safemath{\subD}{\matS} %subdictionary
\safemath{\subA}{\matS_a} %subdictionary part of A
\safemath{\subB}{\matS_b} %subdictionary part of B
\safemath{\dicta}{\matA} 	% first subdictionary
\safemath{\dictb}{\matB} 	% second subdictionary
\safemath{\hollowS}{H}
\safemath{\hollowA}{H_a}
\safemath{\hollowB}{H_b}
\safemath{\cross}{Z}
\safemath{\coh}{\mu_d}			% coherence dictionary
\safemath{\coha}{\mu_a}			% coherence first subdictionary
\safemath{\cohb}{\mu_b}			% coherence second subdictionary
\safemath{\mubs}{\nu}	%block sub-coherence
\safemath{\cohm}{\mu_m} %mutual coherence
\safemath{\dictset}{\setD}	% set of dictionaries
\safemath{\dictsetp}{\dictset(\coh,\coha,\cohb)}	% set of dictionaries parametrized
\safemath{\dictsetgen}{\dictset_\text{gen}}
\safemath{\dictsetgenp}{\dictsetgen(\coh)}
\safemath{\dictsetonb}{\dictset_\text{onb}}
\safemath{\dictsetonbp}{\dictsetonb(\coh)}

\safemath{\leftside}{U}
\safemath{\rightsideA}{R_a}
\safemath{\rightsideB}{R_b}

\safemath{\indexS}{\setI_S} %set of indices participating in sub-dictionary S

\safemath{\na}{n_a}			% cardinality of set of linearly independent columns of first dictionary
\safemath{\nb}{n_b}			% cardinality of set of linearly independent columns of second dictionary
\safemath{\coeffa}{p_i}	%coefficients for columns of A
\safemath{\coeffb}{q_j}	%coefficients for columns of B
\safemath{\seta}{\setP}		% set of linearly independent columns of A
\safemath{\setb}{\setQ}     % set of linearly independent columns of B
\safemath{\setw}{\setW}	%set of n largest elements of w
\safemath{\setz}{\setZ}	%set of L-n largest elements of z
\safemath{\cola}{\veca}		% generic element of the dictionary A
\safemath{\colb}{\vecb}		% generic element of the dictionary B
\safemath{\cold}{\vecd}		% generic element of the dictionary D
\safemath{\inputvec}{\vecx} 	%coefficient vector (input)
\safemath{\error}{\vece}	%error vector
\safemath{\noiseout}{\vecz} 	%noisy output vector
\safemath{\inputvecel}{x}
\safemath{\inputveca}{\vecx_a}
\safemath{\inputvecb}{\vecx_b}
\safemath{\outputvec}{\vecy}	%output of Dictionary
\safemath{\lambdamin}{\lambda_{\mathrm{min}}}
\safemath{\elltwo}{\ell_2}
\safemath{\ellone}{\ell_1}
\safemath{\ellzero}{\ell_0}
\safemath{\ellinf}{\ell_\infty}
\safemath{\ellinftilde}{\ell_{\widetilde\infty}}
\safemath{\licard}{Z(\coh,\coha,\cohb)}
\safemath{\xsol}{\hat{x}}
\safemath{\xbord}{x_b}		%Solution at the border
\safemath{\xstat}{x_s}		%Solution stationary in l0 prob
\safemath{\xstatLone}{\tilde{x}_s}
\safemath{\order}{\mathcal{O}} %order notation (big O)
\safemath{\scales}{\Theta} %scales as
\safemath{\ones}{\mathbf{1}} %all ones matrix
\safemath{\zeroes}{\mathbf{0}} %all zeroes matrix
\safemath{\thlone}{\kappa(\coh,\cohb)} %treshold l1 problem
\safemath{\constoneA}{\delta} %constant in l1 theorem to save space
\safemath{\constoneB}{\epsilon} %constant in l1 theorem to save space
\safemath{\nlarge}{L}				   %num large elements
\safemath{\sumlarge}{S_\nlarge}
\safemath{\maxlarger}{P_\nlarge}	   % maximum in Gribonval and Nielsen
\safemath{\Pzero}{\textrm{P0}}	
\safemath{\Pone}{\textrm{P1}}
\safemath{\vecfir}{\vecw}			 % \vecv element of the kernel of the dictionary, \vecv=[\vecfir \vecsec]
\safemath{\vecsec}{\vecz}
\safemath{\elvecfir}{w}              % element of vecfir
\safemath{\elvecsec}{z}				 % element of vecsec
\safemath{\nlargefir}{n}
\safemath{\normout}{\gamma}
\safemath{\auxfun}{h}
\safemath{\supp}{\textrm{supp}}%support

\safemath{\indexa}{\ell}
\safemath{\indexb}{r}
\safemath{\indexc}{i}
\safemath{\indexd}{j}

\safemath{\project}{P}%projector

\newcommand*{\fancyrefasslabelprefix}{ass}
\frefformat{vario}{\fancyrefasslabelprefix}{Assumption~#1}

\IEEEoverridecommandlockouts

\newtheorem{rem}{Remark}

\safemath{\Qm}{Q_\textnormal{m}}
\safemath{\deltat}{{\scriptstyle\Delta}t}
\safemath{\modulo}{\mathbin{\texttt{\%}}}

\safemath{\fd}{f_{\scriptscriptstyle\!\delta}}

\safemath{\fc}{f_\textnormal{c}}
\safemath{\fmax}{f_\textnormal{max}}
\safemath{\virg}{\textnormal{,}}
\safemath{\deltar}{\Delta_\textnormal{r}}
\safemath{\betattd}{\beta_\textnormal{\tiny TTD}^\theta}
\safemath{\betanb}{\beta_\textnormal{\tiny NB}^\theta}
\safemath{\MLL}{\textit{MLL}}
\safemath{\SLL}{\textit{SLL}}
\safemath{\Epsilon}{\mathcal{E}}
\safemath{\sigmar}{\sigma_\textnormal{\!g}}
\safemath{\sigmai}{\sigma_\textnormal{\!p}}
\safemath{\sigmatot}{\sigma_\textnormal{tot}}
\safemath{\sigmat}{\sigma_\textnormal{\!t}}
\DeclareRobustCommand{\PGFMarker}[2]{%
  \tikz[baseline=-0.6ex]{%
    \begingroup
      \pgfplotmarksize=#2\relax
      \pgfuseplotmark{#1}%
    \endgroup
  }%
}

\begin{document}

\title{The Influence of Gain and Phase Mismatches\\on Beam Patterns in Phased Arrays}

\author{J\'er\'emy Guichemerre and Christoph Studer\\[0.3cm]
\thanks{This work was funded in part by the Swiss State Secretariat for Education, Research, and Innovation (SERI) under the SwissChips initiative. This work was also supported in part by the Swiss National Science Foundation (SNSF) through the Project ``Ubiquitous Large InTelligent ArRAys (ULTRA)'' under Grant 219710 and the HORIZON-JU-SNS-2024-STREAM-C ``X-TREME 6'' Project under Grant 101192681.}
\thanks{J.~Guichemerre and C.~Studer are with the Department of Information Technology and Electrical Engineering, ETH Zurich, Switzerland (e-mail: jeremyg@iis.ee.ethz.ch, studer@ethz.ch)}
}

\maketitle
% !TEX root = ../Beam_mismatch.tex
% DO NOT REMOVE THE ABOVE COMMENT!

\begin{abstract}
Practical implementations of phased arrays suffer from per-antenna gain, phase, and delay mismatches, which can significantly worsen the maximum sidelobe level (SLL) of beampatterns.
The existing literature either analyzes specific structured mismatch patterns or derives per-angle marginal statistics under random mismatches, which fail to characterize global beampattern metrics such as the maximum SLL.
To address this limitation, we propose a frequency-domain framework in which the beampattern is described by a tapering-window-dependent base function evaluated along a deformation determined by the array architecture and signal bandwidth.
This formulation enables a spectral analysis of mismatches, revealing that element-wise errors generate weighted replicas of the ideal beampattern whose amplitudes are given by the discrete Fourier transform of the mismatch sequence.
Building on this insight, we derive an approximation of the maximum SLL distribution under random gain and phase mismatches.
The resulting expressions enable yield-oriented design and rapid design-space exploration without relying on computationally intensive Monte--Carlo simulations.

\end{abstract}

% !TEX root = ../Beam_mismatch.tex
% DO NOT REMOVE THE ABOVE COMMENT!

\section{Introduction}

Fifth generation (5G) and beyond-5G wireless systems require additional spectrum to support the ever-increasing data-rate demands.
Spectrum scarcity drives operation toward millimeter-wave (mmWave) frequency bands, but operation at mmWave frequencies introduces severe free-space path loss and increased sensitivity to blockers~\cite{Rappaport13,Swindlehurst14}.
At such carrier frequencies, beamforming  becomes a key enabler to maintain link budget and spatial selectivity.
Low-Earth orbit (LEO) satellite constellations also constitute a major driver for advanced beamforming architectures~\cite{Angeletti25,Laursen23}.
LEO satellite communication (Satcom) systems aim at providing similar data rates as their terrestrial counterparts, also pushing their operations to mmWave frequencies with even higher path loss.
For such systems, beamforming is necessary to achieve usable link budgets while allowing ground terminals to track fast-moving satellites.
Therefore, LEO Satcom systems not only require accurate beamsteering capabilities but also strict sidelobe control over wide bandwidths in order to reduce out-of-angle interference. 
Electronically steerable phased arrays provide the hardware platform that enables such beamforming capabilities in both terrestrial mmWave systems and LEO Satcom systems.
Both use cases require precise control of the array beampattern over a wide angular range and, in broadband operation, over a wide frequency range.

\subsection{Tapering and Mismatches}

In order to avoid transmitting signals in an unwanted direction or to reject spatial blockers (e.g., from out-of-angle interference), array designers commonly shape the sidelobe patterns of an antenna system through amplitude tapering~\cite{Brookner91}.
Tapering window functions reduce the maximum sidelobe level (SLL) at the cost of widened main lobes and reduced peak beamforming gain.
Design procedures therefore target a prescribed SLL while keeping the main-lobe width within system constraints.
Broadband operation further couples the angular and frequency responses~\cite{Rotman16} and complicates the selection of tapering window coefficients.

Practical phased-array implementations suffer from gain and phase mismatches across antenna elements~\cite{Zhang24,Muriel21,Hu26}.
Such gain and phase mismatches perturb the intended tapering coefficients, which affects the beampattern.
Small mismatches typically induce only modest main-beam degradation, but can substantially increase SLLs.
Periodic or structured mismatches may further generate replica lobes whose locations and amplitudes depend on the mismatch pattern.

The literature provides a variety of models to characterize array imperfections~\cite{He21}.
References~\cite{Chen25,Vorobyov03} analyze beam degradation at the target steering angle only; references~\cite{Hsiao85,Biggelaar18} derive per-angle marginal distributions for the array response under random gain or phase errors.
Per-angle marginal statistics, however, do not characterize global beampattern properties, such as the maximum SLL, because the array response exhibits strong angular correlation.
System-level specifications often impose constraints on the worst-case SLL or on the probability that the SLL exceeds a prescribed threshold.
Thus, existing results do not provide a tractable framework that links element-wise mismatches to such global sidelobe metrics that are valid over large signal bandwidths.

\subsection{Contributions}

We develop a framework to analyze the impact of gain and phase mismatches on the beampattern of phased arrays. Our key contributions are as follows.
First, we introduce a frequency-domain representation of phased-array beamforming that expresses the beampattern as the composition of a base function determined by the tapering window and a deformation function capturing the frequency-dependent effects of true-time-delay (TTD) and narrowband (NB) architectures.
Second, we analyze the tradeoff between main-lobe level (MLL) and sidelobe level (SLL) under practical dynamic-range constraints on tapering coefficients. Using convex optimization, we characterize the achievable region for symmetric tapering windows and compare classical window functions against this region.
Third, we derive a spectral model for gain mismatches showing that element-wise mismatches generate weighted replicas of the ideal beampattern whose amplitudes correspond to the discrete Fourier transform (DFT) of the mismatch sequence. This representation provides an intuitive explanation of replica lobes caused by structured mismatch patterns.
Fourth, we develop an approximation of the maximum SLL distribution under random gain and phase mismatches, which enables yield-oriented SLL predictions without extensive Monte–Carlo simulations.
Finally, we demonstrate how the proposed analysis can be used to derive practical design specifications for large phased arrays.

% !TEX root = ../Beam_mismatch.tex
% DO NOT REMOVE THE ABOVE COMMENT!

\subsection{Notation}

We write matrices and vectors in bold uppercase and bold lowercase, respectively. 
We index the entries $u_k$ of a vector $\bmu\in\complexset^N$ with $k$ in $\llbracket0,N\!-\!1\rrbracket$, where we write $\llbracket a,b\rrbracket$ for the set of integers ranging from $a$ to $b$. 
We define the $k$th entry of the discrete Fourier transform (DFT) of a vector $\bmu\in\complexset^N$ as  
\begin{align}
    U_k = \{\bF\bmu\}_k
        =  \frac{1}{N}\sum_{n=0}^{N-1} \mathrm{e}^{-2\pi j \frac{kn}{N}}u_n,
\end{align}
with $\bF$ being the associated DFT matrix and  $j$ the imaginary unit. 
We define the discrete-time Fourier transform (DTFT) of a complex sequence $\{u_n\}_{n\in\integers}$ as
\begin{align} \label{eq:DTFTdef}
	U(\omega) = \sum_{n=-\infty}^{+\infty} u_n \mathrm{e}^{-2\pi j n \omega},
\end{align}
where $\omega\in\reals$ is measured in \emph{turns}; under this convention, the DTFT $U$ is $1$-periodic.
The circular convolution $U\circledast V$ of the functions $U$ and $V$ associated to the DTFT definition in~\fref{eq:DTFTdef} is 
\begin{align}
	\left(U\circledast V\right)(\omega) = \int_{0}^{1} U(\tau)V(\omega-\tau) \mathrm{d}\tau.
\end{align}
We write the complex conjugate of $z\in\complexset$ as~$z^*$ and $n\modulo N$ as a shorthand for $n\!\!\mod N$.
Expectation of a random variable~$X$ is denoted by $\Ex{X}$ and $\ln(\cdot)$ is the natural logarithm.

\subsection{Paper Outline}
The rest of the paper is organized as follows.
\fref{sec:Framework} introduces the frequency-domain framework for beampatterns of uniform linear arrays (ULAs).
\fref{sec:beamtapering} introduces the basics of antenna tapering.
\fref{sec:beammismatchsec} investigates the impact of hardware mismatches on beampatterns. 
\fref{sec:UPAextension} outlines how our results can be extended to uniform planar arrays (UPAs).
\fref{sec:limitations} highlights limitations of our framework. 
\fref{sec:conclusions} concludes.

% !TEX root = ../Beam_mismatch.tex
% DO NOT REMOVE THE ABOVE COMMENT!

\section{Beampattern Analysis Framework} \label{sec:Framework}

In order to analyze the effect that mismatches have on the transmitted signal in the far-field of a uniform linear array (ULA), we first develop a framework to analyze the link between the per-antenna gain and the resulting beampattern.

%% ------------------------------------------------------------------ %%
%% ------------------------------------------------------------------ %%
\subsection{Time-Domain System Model} \label{sec:beamttdmodel}

\begin{figure}[tp]
	\centering
	\includegraphics[width=0.95\linewidth]{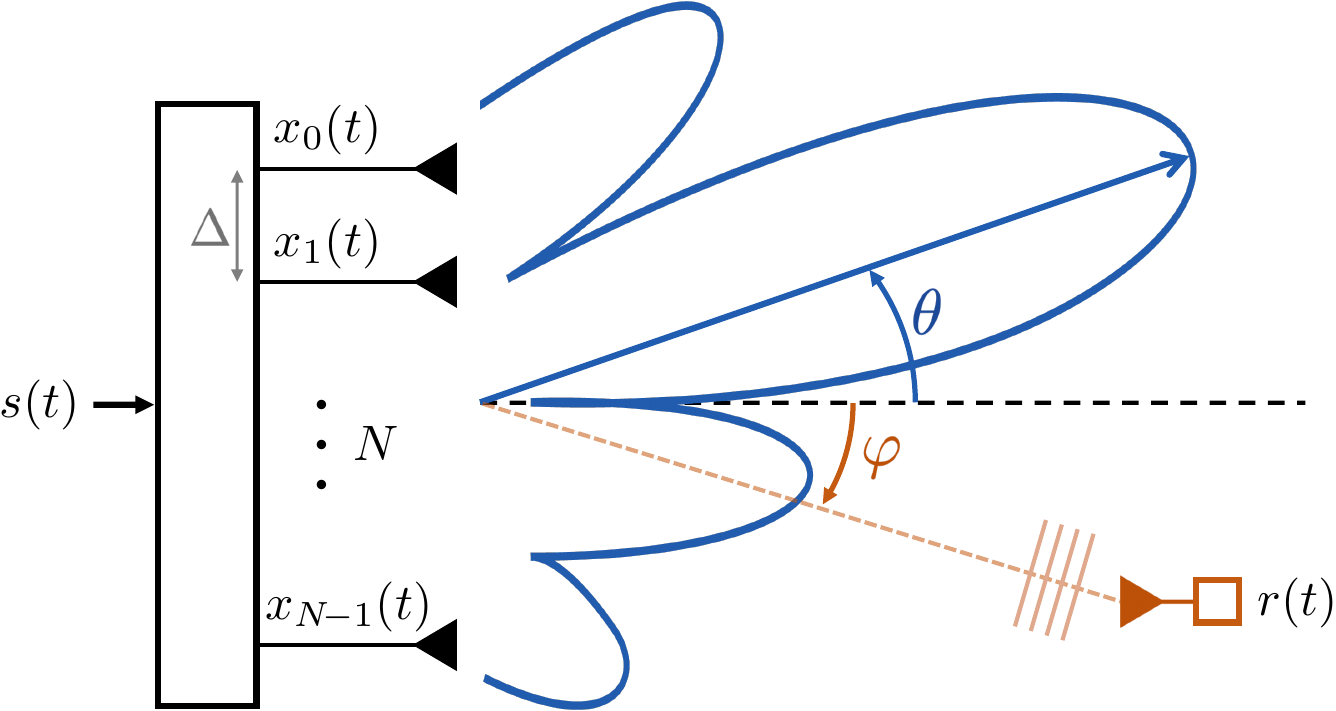}
	\vspace{-0.15cm}  
	\caption{Overview of the considered system. An $N$-antenna ULA with antenna spacing $\Delta$ beamforms the baseband signal $s(t)$ towards the steering angle $\theta$. We observe the effectively radiated signal in far-field at the probing angle $\varphi$, where positive angles are defined counter-clockwise.}
	\label{fig:2_angleSetup}
\end{figure}

We consider a ULA of $N$ antennas with antenna spacing~$\Delta$ (measured in meters).\footnote{A generalization to UPAs is provided in \fref{sec:UPAextension}.}
In what follows, we assume the distance between transmitter (TX) array and receiver (RX) to be significantly larger than the total array size. 
Under this far-field assumption, we can utilize the plane-wave approximation and assume that the signals originating from each antenna experience the same path loss.
For a one-dimensional ULA, we therefore only need to define the \emph{steering angle} $\theta\in[-90^\circ,90^\circ]$ to which we want to focus the transmitted beam and the \emph{probing angle} $\varphi\in[-90^\circ,90^\circ]$ at which we measure how much power is effectively radiated in the far-field; the considered scenario is illustrated in~\fref{fig:2_angleSetup}.

\begin{rem}
We perform our analysis from the point of view of a TX steering its transmit signal to the angle $\theta$, and we observe the effect of the steering in the far-field at the probing angle $\varphi$.
However, all of our results are symmetrically valid for an RX which beamforms to the angle $\theta$ in the presence of a plane-wave reaching the ULA with incident angle $\varphi$.
\end{rem}

When we probe the boresight angle of the array in the far-field, i.e., when ${\varphi=0^\circ}$, each passband transmit signal~$x_{n}^\textnormal{pb}(t)$, $t\in\reals$, $n\in\llbracket0,N-1\rrbracket$, originating from the $n$th antenna, experiences the same time delay from its transmit antenna to the far-field probing point.
However, at any other probing angle~$\varphi$, the delays between a given transmit antenna and the far-field probing point will vary depending on the antenna index.
Without loss of generality, we consider that the transmit signal $x_0^\textnormal{pb}(\cdot)$ at antenna $0$ never experiences delay no matter the angle and define the delays of the other signals with respect to the signal at antenna $0$.

We are interested in the complex baseband  representation of the transmitted and received signal.
We therefore need to translate the impact of a physical delay applied to a passband signal to its baseband representation.
By converting a delayed passband signal (by a delay $\tau$) into baseband, we obtain 
\begin{align} \label{eq:2_delayBB}
	x^\textnormal{pb}(t-\tau) \:\: \overset{BB}{\longrightarrow} \:\: x(t-\tau)\mathrm{e}^{-2\pi j \fc\tau} ,
\end{align} 
where the \emph{baseband transmit signal}~$x(t)$, $t\in\reals$, is the complex baseband representation (centered around $0$\,Hz) of the real passband signal $x^\textnormal{pb}(\cdot)$ (centered around the carrier frequency~$\fc$). 
With this, we can express the complex baseband signal $r$ observed at probing angle $\varphi$ in the far-field as the sum of the influence of each of the $N$ antennas as follows:
\begin{align} \label{eq:2_recombin}
	 r(\varphi, t) = \frac{g(\varphi)}{\ell}	\sum_{n=0}^{N-1}\textstyle x_n\!\left( t - \frac{n\deltar\sin(\varphi)}{\fc}\right) \mathrm{e}^{-2\pi j n\deltar\sin(\varphi)}.
\end{align}
Here, $x_n(\cdot)$, $n\in\llbracket0,N-1\rrbracket$, are the per-antenna baseband transmit signals (associated with the passband transmit signals~$x^\textnormal{pb}_n(\cdot)$,  $n\in\llbracket0,N-1\rrbracket$), $\deltar=\Delta/\lambda$ is the antenna spacing normalized to the wavelength $\lambda$, $g(\varphi)$ is the TX antenna radiation pattern (ARP), and~$\ell$ models the path loss between the TX array and the RX probing direction. 
The signal originating from the $n$th antenna travels an additional distance $n\Delta\sin(\varphi)$ following the plane-wave assumption.

%% ------------------------------------------------------------------ %%
%% ------------------------------------------------------------------ %%
\subsection{True Time-Delay  Beamforming}
\label{sec:TTDbeamformingSICK}

The TX wishes to transmit the \emph{baseband signal} $s(\cdot)$ in the direction specified by the steering angle $\theta$.
In what follows, we use maximum ratio transmission (MRT)~\cite{Lo99}, which maximizes the power in the far-field towards the probing angle $\varphi=\theta$.
MRT counteracts the delay and phase shift each per-antenna transmit signal $x_n(\cdot)$ experiences according to~\fref{eq:2_recombin}.
We refer to the approach presented in this section as \emph{true time-delay} (TTD) \emph{beamforming}, because it compensates both for the time delays and phase shifts.
Concretely, to steer the baseband signal $s(\cdot)$ towards the angle $\theta$, the baseband transmit signals for each antenna $n\in\llbracket0,N-1\rrbracket$ are as follows:
\begin{align} \label{eq:2_TTD_2}
	x_{\textnormal{\tiny TTD,}n}(t) = s\!\left( t + \frac{n\deltar\sin(\theta)}{\fc}\right) \mathrm{e}^{2\pi j n\deltar\sin(\theta)}, \,\, t\in\reals.
\end{align}

\begin{rem}
MRT is oftentimes defined with a normalization factor of $1/\sqrt{N}$ to maintain constant total transmitted power, i.e., the sum of the power emitted by all antennas, constant with varying number of antennas $N$.
Since we focus on the effect of beamforming for a given array topology, we exclude this normalization, and we will obtain the intuitive result that the amplitude at the main lobe peak is proportional to $N$. 
\end{rem}

Plugging the TTD beamforming signal from~\fref{eq:2_TTD_2} into~\fref{eq:2_recombin} leads to the following TX signal radiated in the direction of the probing angle~$\varphi$:
\begin{align} \label{eq:2_recombinTTD}
	&r_\textnormal{\tiny TTD}(\varphi,t) = \notag \\
	 &\,\frac{g(\varphi)}{\ell}\!\sum_{n=0}^{N-1} \!\! \textstyle s\!\left( t \!-\! \frac{n\deltar(\sin(\varphi)-\sin(\theta))}{\fc}\right) \!\mathrm{e}^{-2\pi j n\deltar(\sin(\varphi)-\sin(\theta))} \! .
\end{align}
It follows from \fref{eq:2_recombinTTD} that the maximum power in the probing direction $\varphi$ is reached when $\varphi=\theta$, as the sum constructively adds the baseband signal $s(\cdot)$ $N$ times to create what will be the peak of the main lobe.

\begin{rem}
If an ARPs are considered, then the maximum power angle can slightly deviate from its ideal value $\theta$ due to the main lobe being reshaped by $g(\varphi)$. However, even if the power peak might deviate, then MRT still maximizes the power sent to the steering angle $\theta$.
We also notice that the main lobe shape depends on $\theta$ because of the difference of sine functions, but, in general,~\fref{eq:2_recombinTTD} provides limited intuition on the power present when probing at an angle $\varphi\neq\theta$.
\end{rem}
%

%-------------------

\subsection{Frequency-Domain System Model}

In order to gain more insight into what power is received at other probing angles $\varphi\neq\theta$, we compute the Fourier transform~(FT) of~\fref{eq:2_recombinTTD} with respect to the time variable $t$ and obtain the frequency-domain far-field transmitted signal
\begin{align}
	R_\textnormal{\tiny TTD}(\varphi, \fd) &= \frac{g(\varphi)}{\ell} S(\fd) \sum_{n=0}^{N-1} \mathrm{e}^{-2\pi j \left(1+\frac{\fd}{\fc}\right)n\deltar(\sin(\varphi)-\sin(\theta))} \nonumber \\
								&= \frac{g(\varphi)}{\ell}S(\fd) \tilde{A}_\textnormal{rect}\!\left( \betattd(\varphi, \fd)\right)\!, \label{eq:ttdR}
\end{align}
with the FT $S(\fd)$, $\fd\in\reals$, of $s(t)$, $t\in\reals$, and where we define the TTD \emph{deformation function} as
\begin{align} \label{eq:betattddef}
	\betattd(\varphi,\fd)=\deltar (\sin(\varphi)-\sin(\theta))\!\left(1+ \frac{\fd}{\fc} \right)\!,
\end{align}
and the \emph{base function} as
\begin{align}
	\tilde{A}_\textnormal{rect}(\omega)=  
	\begin{cases}
		N & \textnormal{if } \omega\in\integers, \\
		\frac{\sin\left(\pi N \omega\right)}{\sin\left(\pi \omega\right)} \mathrm{e}^{-\pi j (N-1)\omega} 
		&  \textnormal{otherwise}.
	\end{cases}
\end{align}
As we take the FT of baseband signals, the frequency variable in~\fref{eq:ttdR} and~\fref{eq:betattddef} is the baseband frequency $\fd=f-\fc$, where $f$ is the frequency of the real passband signal.
For example, the baseband DC frequency $\fd=0$\,Hz maps to the carrier frequency~$\fc$ of the real-valued passband signal.

\begin{figure*}[tp]
	\centering
	\subfigure[Base function $\tilde{A}_\textnormal{rect}$]{\includegraphics[width=0.325\linewidth]{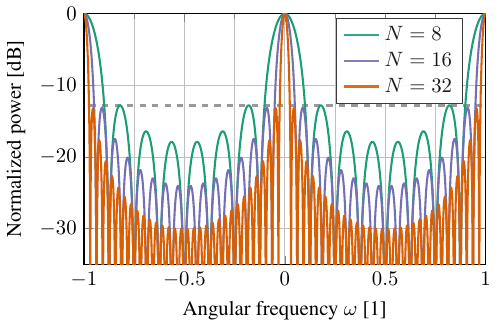}\label{fig:2_baseFunc}}
	\subfigure[TTD beampattern]{\includegraphics[width=0.325\linewidth]{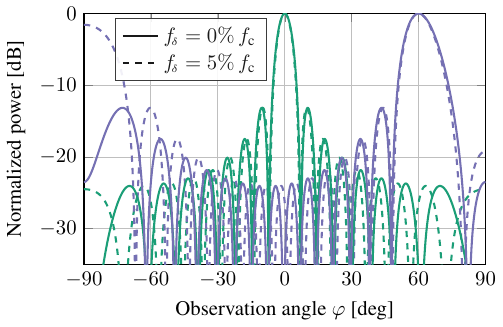}\label{fig:2_TTD}}
	\subfigure[NB beampattern]{\includegraphics[width=0.325\linewidth]{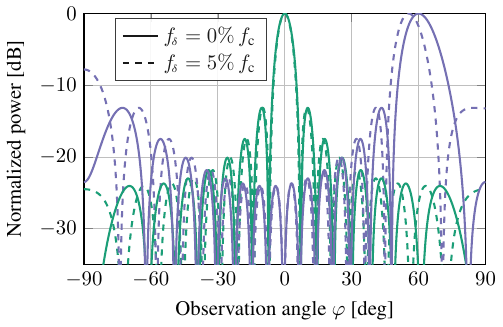}\label{fig:2_NB}}
	\vspace{-0.15cm}  
	\caption{(a) Base function $\tilde{A}_\textnormal{rect}$ for different number of antenna elements $N$; the SLL remains at approx. $13\,$dB in all cases. (b) and (c) show beampatterns for $N=16$ antenna elements with steering angles of $0^\circ$ and $60^\circ$ and two baseband frequencies, in the case of TTD and NB beamforming, respectively; the NB beampattern shows beam squint when $\theta\neq 0^\circ$.}
\end{figure*}

The base function $\tilde{A}_\textnormal{rect}$ is continuous and 1-periodic, and has zeros placed on every multiples of $1/N$ within one period except at 0.
Those zeros lead to a large main lobe of peak power $N^2$ at $\omega=0$, and to $N-2$ sidelobes\footnote{When $N$ is odd, $N-2$ sidelobes are still present, but one of the lobe is split between $-\!\left[\frac{1}{2}, -\frac{1}{2}\!\left(1-\frac{1}{N}\right)\right]$ and $\!\left[\frac{1}{2}\!\left(1-\frac{1}{N}\right), \frac{1}{2}\right]$}.
\fref{fig:2_baseFunc}~shows the base function~$\tilde{A}_\textnormal{rect}$ for different values of $N$.

\subsection{Beampattern}

We now define the beampattern as the ratio between the energy spectral density (ESD) of the receive baseband signal~$|R_\textnormal{\tiny TTD}(\varphi, \fd)|^2$, probed at the probe angle $\varphi$, and the ESD $|S(\fd)|^2$ of the baseband transmit signal $s(t)$. 
In what follows, we ignore the path loss as it does not result from the TX architecture and also the impact of the ARP as it is a fixed $\varphi$-dependent multiplication factor.\footnote{We will briefly discuss the impact of ARPs on our results in~\fref{sec:beamtapering}.}
With those assumptions, the beampattern is defined as follows:
\begin{align}
	P_\textnormal{\tiny TTD}(\varphi, \fd) = \frac{|R_\textnormal{\tiny TTD}(\varphi, \fd)|^2}{|S(\fd)|^2}= 
	\left|\tilde{A}_\textnormal{rect}\!\left( \betattd(\varphi, \fd)\right)\right|^2\!.
\end{align}
We notice that the base function $\tilde{A}$ defines the overall shape of the beampattern, and that the deformation function $\betattd$ defines the \emph{evaluation segment} over which $\tilde{A}$ is evaluated and deformed.
As expected from our discussion about~\fref{eq:2_recombinTTD}, the deformation function $\betattd$ always crosses $0$ when $\varphi=\theta\,$: the main lobe of the base function~$\tilde{A}$ is therefore always centered at $\varphi=\theta$.

For the usual case of $\deltar=0.5$, which corresponds to $\lambda /2$ antenna spacing, the evaluation segment is of length $1+\fd/\fc$.
If $\fd<0$, then the length of the evaluation segment  is smaller than one, meaning that the beampattern contains at most one period of $\tilde{A}$, and, therefore, contains at most $N-2$ sidelobes.
However, if $\fd>0$, i.e., if $f>f_c$, then more than one period of $\tilde{A}$ is mapped to the beampattern, which can lead to more than $N-2$ sidelobes in the beampattern.
In cases where either~$\deltar$ exceeds 0.5 or where $\fd$ is positive, the main lobe itself can get replicated; such a replica of the main lobe is called a grating lobe.
The further the steering angle $\theta$ deviates from boresight ($0^\circ$), the closer a grating lobe is from the explored segment boundary. This effect is visible in~\fref{fig:2_TTD}, in which targeting a steering angle of $\theta=60^\circ$ already leads to a partial grating lobe at $\varphi=-90^\circ$ for a baseband frequency $\fd=0.05\fc$.

The deformation function $\betattd$ not only changes which portion of the base function gets mapped to the beampattern, but also stretches the base function by the $\sin(\varphi)$ component.
As the derivative of $\sin(\varphi)$ is close to one for small values of~$\varphi$, the deformation function $\betattd$ will not have a significant impact on the main lobe at boresight. However, the more $\varphi$ deviates from zero, the smaller the derivative of $\betattd$, which leads to $\tilde{A}$ being more and more stretched.
This effect leads to the main lobe becoming wider when the steering angle $\theta$ deviates from $0^\circ$, as can be seen in~\fref{fig:2_TTD}.\footnote{As discussed in \fref{sec:TTDbeamformingSICK}, the ARP also affects the beampattern as a fixed $\varphi$-dependent multiplication factor.}

%% ------------------------------------------------------------------ %%
%% ------------------------------------------------------------------ %%
\subsection{Narrowband Beamforming}
  
We showed previously that TTD beamforming allows to aim at the correct steering angle no matter the frequency of the signal to be transmitted.
However, TTD beamforming requires tunable time-delay circuitry, which is difficult (or costly) to implement in hardware. 
For this reason, in the majority of contemporary hardware implementations, the time delay in~\fref{eq:2_TTD_2} is ignored and only the phase shift is considered, leading to
\begin{align} \label{eq:2_NB}
	x_{\textnormal{\tiny NB,}n}(t) = s\!\left( t \right) \mathrm{e}^{2\pi j n\deltar\sin(\theta)}, \,\, t\in\reals,
\end{align}
for a steering angle $\theta$. 
We refer to this case as \emph{NB beamforming}; the reason for this name becomes apparent when we apply the same procedure as in the previous section. We obtain
\begin{align}
	P_\textnormal{\tiny NB}(\varphi, \fd) = \frac{|R_\textnormal{\tiny NB}(\varphi, \fd)|^2}{|S(\fd)|^2}= |\tilde{A}_\textnormal{rect}\!\left( \betanb(\varphi, \fd)\right)|^2,
\end{align}
with $\betanb$ the \emph{NB deformation function} defined as
\begin{align}
	\betanb(\varphi,\fd)=\deltar \!\left(\sin(\varphi)\!\left(1+ \frac{\fd}{\fc} \right)-\sin(\theta)\right).
\end{align}
We notice that the base function did not change, which means that the overall shape of the beampattern will not be impacted.
However, the distortion function $\betanb$ differs from $\betattd$ in that the frequency-dependent factor $1+\fd/\fc$ is only multiplied to $\sin(\varphi)$.
In practice, as the zero of $\betanb$ does not occur for $\varphi=\theta$ anymore (except for a boresight steering angle $\theta=0^\circ$), the main lobe of the beam pattern deviates from its nominal angle $\theta$ depending on the signal frequency. 
This effect is known as \emph{beamsquint}.
In fact, solving for the zero of $\betanb$, we obtain  
\begin{align}
	\varphi_\textnormal{main} = \arcsin\!\left(\frac{\fc}{\fd + \fc} \sin(\theta)\! \right),
\end{align}
where $\varphi_\textnormal{main}$ is the probing angle at which the peak of the main lobe effectively occurs for a given baseband frequency~$\fd$.
Beamsquint occurs in the context of NB beamforming but not of TTD beamforming because the phase shift in \fref{eq:2_delayBB} on the baseband signal has been taken into account, but not the time delay.
In practice, the receiver only probes one angle $\varphi$; beam squint therefore acts as a frequency filter, potentially leading to increased inter-symbol interference and/or reduced signal to noise ratio.

% !TEX root = ../Beam_mismatch.tex
% DO NOT REMOVE THE ABOVE COMMENT!

\section{Tapering} \label{sec:beamtapering}

The beamforming methods introduced above are optimal in the sense that the energy transmitted  to the steering angle, i.e., the main-lobe level (MLL), is maximized.
However, another practically important metric is the sidelobe level (SLL) which quantifies the power of the strongest secondary lobe relative to the main one.
From a TX perspective, SLL is important not to disrupt communication outside of the steering angle, e.g., energy radiated towards other satellites.
From an RX perspective, SLL  represents the rejection of signals originating from non-targeted directions.
As we observe from~\fref{fig:2_baseFunc}, methods in~\fref{sec:Framework} reach an SLL of only approximately $13\,$dB: achieving higher SLL (at the cost of reduced MLL) requires what is known as tapering.
We next summarize the concept of tapering and then characterize different tapering windows.

%% ------------------------------------------------------------------ %%
%% ------------------------------------------------------------------ %%
\subsection{Introduction to Tapering} \label{sec:3_1}

Tapering refers to scaling the $n$th antenna output by a gain factor $a_n\in[0,1]$, called the tapering coefficients, and corresponds to windowing in the spatial domain.
The scaling range is limited to $[0,1]$ because practical power amplifiers (PAs) have a limited output power.
We follow the same steps as in \fref{sec:Framework} and, after applying the FT, we obtain the frequency-domain far-field transmitted signal %
\begin{align} \label{eq:3_Rintermediate_amp}
    R_\bma(\varphi, \fd) = S(\fd) 
    \underbrace{\sum_{n=0}^{N-1}  a_n\mathrm{e}^{-2\pi jn \beta^\theta\!(\varphi, \fd) }}_{=\,C(\varphi, \fd)}\!,
\end{align}
where $\beta^\theta$ can be replaced by either $\betattd$ or $\betanb$ depending on which beamforming type is considered.

Let $\tilde{\bma}$ be an extension of \bma defined on $\reals^\integers$ as
\begin{align} \label{eq:3_a_ext}
    \tilde{a}_n=
    \begin{cases}
        a_n \:&\textnormal{if } n\in\llbracket 0,N-1\rrbracket \\
        0 &\textnormal{otherwise,}
    \end{cases}
\end{align}
we can then write
\begin{align} \label{eq:3_Rintermediate_amp_X0}
    C(\varphi, \fd) = \sum_{n=-\infty}^\infty  \tilde{a}_n \mathrm{e}^{-2\pi j n\beta^\theta\!(\varphi, \fd) } = 
    \tilde{A}(\beta^\theta\!(\varphi, \fd)),
\end{align}
with $\tilde{A}(\beta^\theta\!(\varphi, \fd))$ the DTFT of $\tilde{\bma}$ evaluated at $\beta^\theta\!(\varphi, \fd)$.
Note that we implicitly consider that the spatial sampling period of~$\tilde{\bma}$ is $1$ (dimensionless); the effect of the antenna spacing $\Delta$ is already incorporated in $\beta^\theta$ thanks to its $\deltar$ factor.

Substituting $C(\varphi, \fd)$ in \fref{eq:3_Rintermediate_amp} by \fref{eq:3_Rintermediate_amp_X0}, we obtain
\begin{align} \label{eq:3_R_final}
    R_\bma(\varphi, \fd) = S(\fd)  \tilde{A}(\beta^\theta\!(\varphi, \fd)).
\end{align}
The beampattern is therefore given by
\begin{align} \label{eq:3_P_final}
P_{\!\bma}(\varphi, \fd) = |\tilde{A}\!\left( \beta^\theta\!(\varphi, \fd)\right)|^2.
\end{align}

We notice that the results in \fref{sec:Framework} are a special case with $a_n=1$, $n\in\llbracket 0,N-1\rrbracket $, leading to a base function corresponding to the DTFT of a rectangular window.
In general, tapering only affects the base function. The deformation applied to the window-dependent base function does not depend on the window, but only on whether TTD or NB beamforming is employed.

%% ------------------------------------------------------------------ %%
%% ------------------------------------------------------------------ %%
\subsection{Measuring MLL and SLL} \label{sec:method}

We now formally introduce two additional measures that are used when analyzing the properties of beampatterns.

The MLL, which characterizes the loss in transmit power due to tapering, is defined as follows:
\begin{align}
	\MLL(\bma) \triangleq  |\tilde{A}(0)|= \|\bma\|_1 \le N.
\end{align}
Here,  \bma is the vector containing the $N$ tapering coefficients $a_n$, $n\in\llbracket0,N-1\rrbracket$, and $\tilde{A}$ the DTFT of $\tilde\bma$, which is the extension of $\bma$ as defined in \fref{eq:3_a_ext}. 
We stress again that the tapering coefficients are bounded from above by $1$ as the transmit power at each antenna is in practice limited by its corresponding power amplifier.

\begin{rem}
	We implicitly assume that the maximum of the base function is centered at zero (and at all integer positions due to the 1-periodicity). This property emerges naturally from symmetric windows which are the vast majority of typically considered tapering windows. Note that exception do exist, for example from the radar community with the Bayliss window~\cite{Bayliss68}, but we will not be considering those cases here.
\end{rem}

The SLL, which characterizes the ratio between the MLL and the strongest secondary lobe peak, is defined as follows: 
\begin{align} \label{eq:3_SLLdef}
	\SLL(\bma) \triangleq \frac{\MLL(\bma)}{ \|\tilde{A}(\Omega_\textnormal{out})\|_\infty}.
\end{align}
Here, $\Omega_\textnormal{out}$ is the interval in $[-0.5,0.5]$ from which the main lobe support has been excluded.
We note that this definition of $\Omega_\textnormal{out}$ implicitly considers that the entire base function, i.e., an entire period $[-0.5,0.5]$, is relevant to our analysis.\footnote{In some specific cases, e.g., when the antenna spacing is significantly smaller than $\lambda/2$, portions of $\tilde{A}$ far from the main lobe could be ignored.}

As mentioned earlier, the rectangular window, i.e., the case where no tapering is applied, exhibits an SLL of approximately $13$\,dB (see~\fref{fig:2_baseFunc}).
In order to obtain a larger SLL, tapering has to be employed, necessarily leading to a decreased MLL. Therefore, there exists a fundamental tradeoff between MLL and SLL, which implies the existence of an MLL optimum window for a given target SLL. 

\begin{rem} \label{rem:arpforwindow}
If ARPs are taken into account, then MLL and SLL become steering-angle dependent. Indeed, if one targets an angle that does not correspond to the ARP maximum, then side lobes limiting the SLL can be amplified with respect to the main lobe.
In this case, one can consider a single tapering window no matter what the steering angle is: the dynamic range of the ARP within the scan angle can be added to the target SLL. Alternatively, the tapering window can be steering-angle dependent; our analysis then holds at each considered fixed~angle.
\end{rem}

One typically also uses a third measure: the half-power beam width (HPBW). This measure characterizes the angular width of the main lobe from where the power has dropped by $3$\,dB from its maximum.
The HPBW can be interpreted as the capability of the transmitter to focus the power in only the wanted direction, and is known to be approximately proportional to $1/N$~\cite{Plotkin73}. Due to the effect of the deformation function $\beta^\theta$ when the steering angle deviates from boresight, the HPBW also increases with the absolute value of the steering angle $|\theta|$.
In what follows, we do not focus on the HPBW as it is typically less of a concern for large phased arrays, but rather on the MLL vs. SLL tradeoff. Note that, as we will see in~\fref{sec:3_windows}, windows performing well in terms of the MLL vs. SLL tradeoff are also performing well in terms of HPBW.

\begin{figure*}[tp] 
\centering
\subfigure[SLL vs MLL for $N\!=16$ antennas.]{%
	\includegraphics[width=0.48\linewidth]{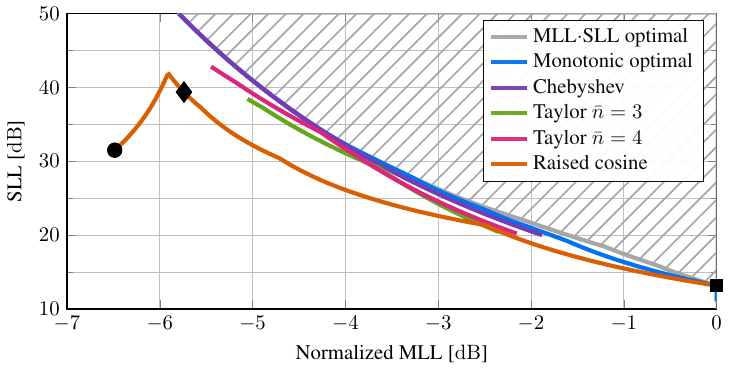}
	\vspace{-0.15cm} 
	\label{fig:3_SLLplane16}} 
\hspace{0.2cm}
\subfigure[SLL vs MLL for $N\!=64$ antennas.]{%
	\includegraphics[width=0.48\linewidth]{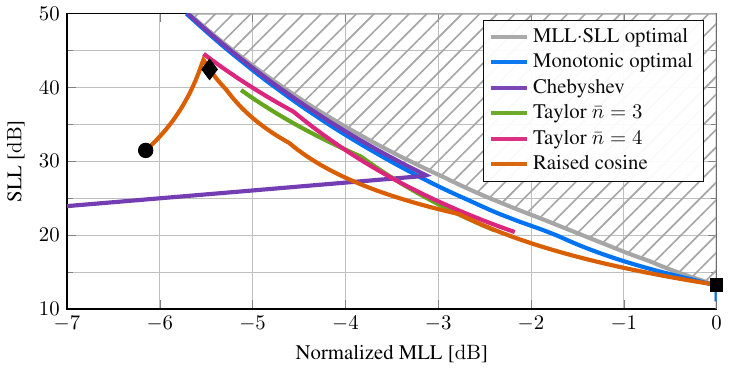}
	\label{fig:3_SLLplane64}}
\vspace{-0.4cm}  
\caption{MLL normalized by the maximum achievable gain and SLL for Chebyshev, raised-cosine, and Taylor windows. Special cases of the raised-cosine windows are marked: \PGFMarker{square*}{2.3pt} rectangular window ($\alpha=1$), \PGFMarker{diamond*}{3pt} Hamming window ($\alpha\approx 0.54$), and \PGFMarker{*}{2.3pt} Hann window ($\alpha=0.5$). The optimal achievable MLL for a given SLL is also shown and the unachievable area of MLL/SLL pairs are hatched; monotonic optimal windows only incur a minor MLL loss.} \label{fig:3_SLLplanes}
\end{figure*}

\begin{figure*}[tp]
	\centering
	\subfigure[Chebyshev windows, $N\!=16$ antennas]{\includegraphics[width=0.325\linewidth]{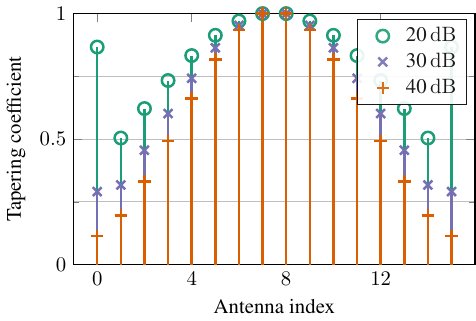}\label{fig:3_cheby16}}
	\subfigure[Chebyshev windows, $N\!=64$ antennas]{\includegraphics[width=0.325\linewidth]{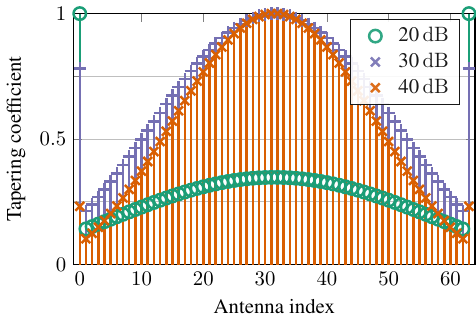}\label{fig:3_cheby64}}
	\subfigure[Monotonic optimal windows, $N\!=64$ antennas]{\includegraphics[width=0.325\linewidth]{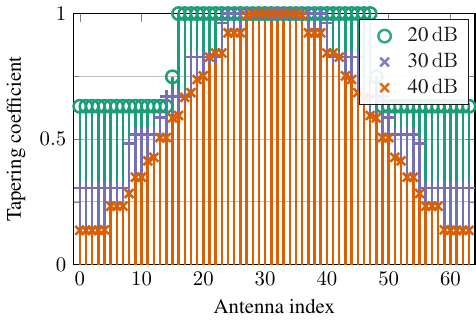}\label{fig:3_optmono64}}
	\vspace{-0.15cm}  
	\caption{(a), (b) Chebyshev and monotonic MLL-SLL optimal window coefficients for various SLL targets (the values in the legends), illustrating the window-edge peaking of Chebyshev windows when the SLL decreases; coefficient peaking worsens at a fixed SLL target when the number of antennas $N\!$ increases. (c) Coefficients of monotonic optimal windows which avoid edge peaking at a minor MLL loss.}
\end{figure*}

%% ------------------------------------------------------------------ %%
%% ------------------------------------------------------------------ %%
\subsection{MLL vs. SLL Tradeoff} \label{sec:tradeoffMLLSLL}

In order to compare different tapering windows, one can assess their MLL for a given SLL target.
Typically, increasing the SLL would come at the cost of reduced MLL. 
A useful visualization of this tradeoff is to plot each tapering window as a point in the normalized MLL/SLL plane, where $0$\,dB normalized MLL corresponds to the MLL of the non-tapered case (i.e., a rectangular window).
Before discussing the efficiency in terms of the MLL vs.\ SLL tradeoff of typically employed windows, we propose to numerically compute the MLL-optimum window for a given SLL target, similar\footnote{Reference \cite{Lebret97} proposes to use convex optimization to obtain beamforming weights and solves a different convex optimization problem than in \fref{eq:3_problemwecannotsolve}.} to the method in~\cite{Lebret97}.

Ideally, we would like to solve the following problem
\begin{align}
	\argmax_{\bma\in[0,1]^N}  \MLL(\bma) \:\:\:\textnormal{subject to}\:\:\SLL(\bma) \geq l,
\end{align}
where $l$ is the desired target SLL. 
This problem is convex, which becomes evident by rewriting it as follows:
\begin{align} \label{eq:3_problemwecannotsolve}
	\argmax_{\bma\in[0,1]^N}  \sum_{n=0}^{N-1} a_n \:\:\:\textnormal{subject to}\:\:   \|\tilde{A}(\Omega_\textnormal{out})\|_\infty \leq \frac{1}{l}\sum_{n=0}^{N-1} a_n.
\end{align}
Here, we note that a sum is linear, hence also concave. 
With the formulation in \fref{eq:3_problemwecannotsolve}, the only obstacle from using numerical solvers is the continuous nature of the infinity-norm constraint. 
We therefore discretize the frequency axis $\Omega_\textnormal{out}$ to a fine grid.
Since \bma is real, its DTFT is conjugate symmetric. Thus, we can only consider nonnegative frequencies and approximate~$\Omega_\textnormal{out}$ as a vector of length $M$ as follows: 
\begin{align}
	\Omega_\textnormal{out}\approx \boldsymbol{\omega}_\textnormal{out} =[\omega_0, \omega_0+\omega_\textnormal{step},  \omega_0+2\omega_\textnormal{step}, \dots, 0.5].
\end{align}
Here,  $\omega_0$ is close to the positive zero of the main lobe and~$\omega_\textnormal{step}$ the frequency grid step size.\footnote{The choice of a uniform grid is arbitrary; in practice, any grid that samples in close proximity to all secondary lobe peak would yield the same result.}
The value for $\omega_0$ needs to be empirically tuned, but has no impact on the result as long as it is between the main and second lobe, and at a frequency where the window power is not limiting the SLL.

In order to show the boundary of achievable MLLs at given SLLs, we solve the following convex optimization problem using CVX~\cite{cvx,gb08} for different target SLL values
\begin{align} \label{eq:3_cvx_final}
	\argmax_{\bma\in[0,1]^N}  \sum_{n=0}^{N-1} a_n \:\:\:\textnormal{subject to}\:\:   \|\bF_\textnormal{out}\bma\|_\infty \leq \frac{1}{l} \sum_{n=0}^{N-1} a_n,
\end{align}
where \bF is the $M\times N$ discretized DTFT matrix with coefficients given by
\begin{align}
	F_{m, n} = \mathrm{e}^{-2\pi jn\cdot\omega_{\textnormal{out},m}}.
\end{align}

\fref{fig:3_SLLplanes} shows, for the cases of $N=16$ and $N=64$ antennas, the MLL vs. SLL tradeoff boundary, which is the achievable region of windowing functions.

\begin{rem}
By symmetry of the problem in \fref{eq:3_cvx_final}, the resulting MLL vs.\ SLL tradeoff optimal windows are symmetric.
Enforcing this symmetry as a constraint in the optimization problem therefore speeds up the numerical solving by reducing the number of unknowns without affecting its solution.
\end{rem}

%% ------------------------------------------------------------------ %%
%% ------------------------------------------------------------------ %%
\subsection{Commonly Employed Windows} \label{sec:3_windows}

A commonly used window for tapering is the Chebyshev window, also known as the Dolph-Chebyshev window, proposed by Dolph in the context of beamforming in 1946~\cite{Dolph46,Riblet47}.
The window optimizes the beamwidth for a given SLL\footnote{In the context of Chebyshev windows, the optimized beamwidth is not the HPBW but the width between the two zeros defining the main lobe.} and leads to the well-known equiripple solution.
The MLL achieved as a function of the target SLL with 16 antennas is shown in~ \fref{fig:3_SLLplane16}.
We observe that Chebyshev windows are very close to be optimal in terms of the MLL-SLL tradeoff on top of being optimal in terms of beamwidth: it therefore seems like such window functions are the preferred option.

However, because Chebyshev windows are solutions of a min-max spectral optimization problem, the resulting tapering coefficients are not, in general, guaranteed to be compatible with practical hardware implementation constraints.
Indeed, when the target SLL decreases for a fixed number of antennas, the first and last tap of the associated Chebyshev window can exhibit peaky behavior.
\fref{fig:3_cheby16} and \fref{fig:3_cheby64} illustrate this effect by showing Chebyshev window coefficients for different SLL targets with $N=16$ and $N=64$ antennas.
This ``discontinuity'' becomes larger at a fixed SLL when the number of antennas grows---up to the point where the edge coefficients can take values larger than the center of the window, effectively limiting the window MLL. This effect is clearly visible in~\fref{fig:3_cheby64} in the case of a target SLL of $20$\,dB (green stem plot).
When the edge of the window is limiting the MLL, the effect becomes apparent on the MLL-SLL plane, as shown for the case of $N\!=64$ antennas in~\fref{fig:3_SLLplane64} when the  target SLL is lower than $28\,$dB. Note that a peaking of more than $6\,$dB relative to the previous tap is already present at the edge of the window for a target SLL as large as $40$\,dB, see~\fref{fig:3_cheby64}.

In order to address such practical implementation issues, Taylor proposed a family of continuous windows~\cite{Taylor55}, later used as discrete windows known as Taylor windows. 
They are parametrized by a target SLL, but also by a factor $\bar{n}$, leading to $\bar{n}-1$ nearly-constant side lobes on each side of the main lobe.
Unlike the case of Chebyshev windows, letting the side lobes decay leads to a non-minimal beamwidth, but, subject to a good choice of~$\bar{n}$, allows to avoid peaking at the edge of the tapering window.
The choice of $\bar{n}$ is important because, if chosen too low, then the SLL target will not be met\footnote{This effect is visible in~\fref{fig:3_SLLplanes}: indeed, the SLL target for Taylor windows has been swept up to 50dB, but the effectively achieved SLLs are limited by the values chosen for $\bar{n}$.}, but if chosen too high, the same implementability issues as with Chebyshev windows may arise.
In practice, one often chooses $\bar{n}$ slightly larger than its minimum to obtain decaying side lobes and implementation-friendly coefficients at the cost of a small increase in beamwidth~\cite{Jakowatz96}.
As illustrated in \fref{fig:3_SLLplane64}, Taylor windows offer realistic implementations at the cost of an MLL loss in the order of $0.5$\,dB, making them a common choice for phased array tapering.

Note that our problem formulation in \fref{eq:3_cvx_final} also ignores such practical implementation issues. 
However, we can impose a convex constraint that ensures the tapering coefficients to not decrease from index $0$ to $N/2-1$ (for an even $N$), in addition to window symmetry. 
By solving this modified optimization problem, we have a practical method to compute \emph{monotonic optimal windows} (for a given SLL), which resolve the peaking issues of Chebyshev windows but, unlike Taylor windows, in an MLL-optimal manner. 
As shown in \fref{fig:3_SLLplanes}, such monotonic optimal windows come at a negligible MLL loss when compared to their unconstrained MLL-SLL optimal counterparts. Examples of the tapering coefficients of such a monotonic optimal windows are shown in~\fref{fig:3_optmono64}.

Another common type of tapering windows is the family of raised-cosine windows, also known as cosine-on-a-pedestal windows, and consists of a linear combination of a rectangular window (the pedestal) and one period of a cosine~\cite{Harris78}. For an even $\!N$, such windows are defined as
\begin{align}
	a_n^\textnormal{rc} = \frac{1}{K}\!\left(\alpha - (1-\alpha)\cos \!\left(\frac{2\pi n}{N-1} \right)\!\right)\!,
\end{align}
for $n\in\llbracket 0,N\!-1\rrbracket$, where $\alpha\in[1/2,1]$ is the cosine weight controlling the contribution of the pedestal and cosine terms. When $\alpha=1$, the resulting window is a pure rectangular window, and for $\alpha=0.5$, the window is a pure cosine, also known as the Hann window\footnote{Note that the Hann window reaches zero at its edge: one can therefore design the window for $N+2$ antennas and ignore the two zero terms.}.
As $N\!$ is even, the discrete sampling of the window does not include the maximum of the cosine function; the window is explicitly normalized to reach~$1$ at its two central taps with the factor $K$ defined as
\begin{align}
	K =  \left(\alpha - (1-\alpha)\cos \!\left(\frac{\pi N}{N\!-1} \right)\!\right).
\end{align} 

\fref{fig:3_SLLplane16} and \fref{fig:3_SLLplane64} show how raised-cosine windows perform on the MLL-SLL plane for the entire range of $\alpha$. As those windows are only defined through a smoothness heuristic, they do not achieve any kind of optimality.
In fact, they do not perform as well as Taylor or monotonic optimal windows in terms of achievable MLL for a given SLL, and typically also feature wider beamwidth; their use is therefore limited to cases where only a low SLL is desired.

\begin{rem}
A variety of possible windows exist and discussing them is outside the scope of this paper.
We recommend the classic paper of Harris~\cite{Harris78} for more details. 
\end{rem}

% !TEX root = ../Beam_mismatch.tex
% DO NOT REMOVE THE ABOVE COMMENT!

\section{Mismatch} \label{sec:beammismatchsec}

Whether tapering is employed or not, matching between physically different electrical components cannot be perfect. Therefore, analog circuit designers  need to derive specifications based on system requirements in order to know how much uncertainty in the fabricated components can be tolerated.
While calibration schemes can drastically reduce the impact of such mismatches, deriving the maximum calibration step size for a given quantity is still required.

%% ------------------------------------------------------------------ %%
%% ------------------------------------------------------------------ %%
\subsection{Origin of Mismatches}

Per-antenna mismatches in phased-arrays originate from a variety of sources.
At the printed circuit board (PCB) level, unbalanced routing to or from the different antennas leads to both delay and gain mismatches~\cite{Muriel21}; on-chip circuitry, especially low-noise amplifiers (LNAs) in RXs and PAs in TXs, also contribute to the latter~\cite{Zhang24,Hu26}.
In most practical implementations, the phase and time shifts used to beamform are in discrete steps, which also leads to mismatch when the applied values are compared to the ideal ones~\cite{Gray85, Holm92}. In the case of analog or hybrid beamforming, the steps sizes themselves are non-ideal as the circuitry responsible for time delays, phase shifts, and summation are realized with analog circuitry.

%% ------------------------------------------------------------------ %%
%% ------------------------------------------------------------------ %%
\subsection{Effect of Gain Mismatch} \label{sec:4_gain}

In what follows, we model per-antenna gain mismatch as an antenna-dependent scaling factor multiplied to the intended tapering window
\begin{align} \label{eq:4_def}
    \tilde{a}_n= \tilde{m}_n\tilde{w}_n =
    \begin{cases}
        m_nw_n \:&\textnormal{if } n\in\llbracket 0,N-1\rrbracket \\
        0 &\textnormal{otherwise},
    \end{cases}
\end{align}
where $\bmw\in[0,1]^N$ is the intended window and $\bmm\in\reals^N$ the mismatched amplitude sequence. We further define
\begin{align} \label{eq:4_mismatchsequence}
	e_n \triangleq m_n - 1,
\end{align}
the relative mismatch at the $n\,$th antenna.
Without losing generality, we assume that only $\bmw$ is extended by zero padding as in~\fref{eq:3_a_ext}, leaving a free choice for the extension of \bmm as values outside $ \llbracket 0,N-1\rrbracket$ are multiplied by zero; see \fref{eq:4_def}.
In order to gain insight on the impact of mismatches, we extend the mismatched amplitude sequence~$\bmm$ periodically as
\begin{align}
    \tilde{m}_n= m_{n \modulo N}.
\end{align}
Therefore, $\tilde{M}$, the DTFT of $\tilde{\bmm}$, consists of Dirac delta functions placed at multiples of $1/N$ whose amplitudes are determined by the DFT of \bmm.
 Finally, as a point-wise multiplication translates into a circular convolution in the DTFT domain, the mismatched beampattern is
\begin{align} \label{eq:R_final_conv}
    P_{\bmw,\bmm}(\varphi, \fd)  =
   \left| \tilde{A}_\bmm (\beta^\theta\!(\varphi, \fd))\right|^2,
\end{align}
where $\tilde{A}_\bmm\triangleq \tilde{M}\circledast\tilde{W}$ is the mismatched base function.

As the DTFT of the periodic extension $\tilde{\bmm}$ of $\bmm$ is a periodic sequence of Dirac pulses, we can further write 
\begin{align} \label{eq:4_replicas}
	\tilde{A}_\bmm(u) = \tilde{W}(u) + \underbrace{\sum_{n=0}^{N-1} r_n  \tilde{W}\!\left(u-\frac{n}{N}\right)}_{\triangleq\,\Epsilon_\bmm(u)},
\end{align}
where $\bmr=\bF\bme$ is the DFT of the relative mismatch sequence~\bme in \fref{eq:4_mismatchsequence}.
We conclude that the impact of gain mismatch on the base function are replicas of the original base function placed at multiples of~$1/N$ whose amplitudes are scaled by the DFT of the mismatch sequence.
We note that the sum in \fref{eq:4_replicas} is complex; at a given $u$, the phases of each element in $\Epsilon_\bmm(u)$ will impact the resulting beampattern.

\begin{rem}
One implication of~\fref{eq:4_replicas} is that a mismatch sequence which is periodic over the array 	will create strong replicas of the main lobe: Indeed, the power of the mismatch will be located on only one or two DFT bins\footnote{The power will be located in two bins, unless the period is $2$, in which case only the Nyquist bin contains the power of the mismatch.} in \bmr.
This effect is well known in the context of phase quantization errors~\cite{Gray85, Holm92} which lead to strong replicas of the main beam at specific target angles that makes the quantization error periodic. Our formulation in \fref{eq:4_replicas} makes this effect explicit. 
\end{rem}

%% ------------------------------------------------------------------ %%
%% ------------------------------------------------------------------ %%
\subsection{Phase and Delay Mismatch} \label{sec:4_phase}

In the case of phase and delay mismatches, we discriminate two scenarios.
On the one hand, the mismatch can originate from a non-ideality in the complex phase shift applied at baseband; this case is strictly a phase shift.
On the other hand, a delay mismatch at RF can be approximated as a phase shift under a NB assumption.

If the mismatch originates from a phase shift applied at baseband, it can directly be modeled as an antenna-dependent complex-valued phase shift similar to the case in~\fref{eq:4_def} with
\begin{align} \label{eq:4_phaseapprox}
	a_n = w_n \mathrm{e}^{j p_n} \approx w_n(1+jp_n),
\end{align}
where \bmp is the vector containing the phase errors and the right-hand side is the small-angle approximation.
The impact on the beampattern is therefore similar to the one described in~\fref{sec:4_gain}, but with a complex-valued mismatch sequence.

We now consider the impact of a sequence of antenna-dependent delays $\mathbf{\deltat}\in\reals^N$, with each entry $\Delta t_n$ corresponding to a time delay at the $n$th antenna applied to the RF signals in passband. In this case, the impact on the baseband signals is modeled as in~\fref{eq:2_delayBB}, leading to the following result:
\begin{align} 
    R_{\bmw,\mathbf{\deltat}}(\varphi, \fd) = S(\fd) 
    \sum_{n=0}^{N-1}  w_n \mathrm{e}^{-2\pi j(\fc+\fd)\deltat_n} \, \mathrm{e}^{-2\pi jn \beta^\theta\!(\varphi, \fd) }.
\end{align}
Here, we notice that the error in phase multiplied to $w_n$ depends on $\fc+\fd$; in this term, $\fc$ reflects the up-conversion phase misalignment due to the time shift which is independent of the nature of the baseband signal $s$ to be transmitted.
The term~$\fd$ represents the time shift on the baseband signal $s$ itself. At $\fd=0$\,Hz, which corresponds to the DC component of $s$, time shifting has no effect; as the baseband frequency $\fd$ increases, the error caused by time shifts increases.

For moderate mismatches that still allow the beamformer to operate properly, we can assume $\left|(\fc+\fd)\deltat_n\right|\ll1$ for all $n$, allowing a similar approximation as in~\fref{eq:4_phaseapprox}
\begin{align} \label{eq:4_timeapprox}
 	w_n \mathrm{e}^{-2\pi j(\fc+\fd)\deltat_n} \approx w_n(1-2\pi j(\fc+\fd)\deltat_n),
\end{align}
where, unlike all previous cases, we are not able to fully disentangle a base function from the rest of the parameters due to the dependency on \fd.

One possibility to circumvent the problem is to define  {{\fd}-dependent}~base functions instead.
This approach can be used directly on the left-hand side of~\fref{eq:4_timeapprox} and allows one to evaluate the impact of a specific sequence $\mathbf{\deltat}$ for different values of~\fd.
Alternatively one can make a NB approximation, ignore the impact of~\fd in~\fref{eq:4_timeapprox}, and approximate
\begin{align} \label{eq:4_timeapproxNB}
 	w_n \mathrm{e}^{-2\pi j(\fc+\fd)\deltat_n} \approx w_n(1-2\pi j(\fc+f_{{\scriptscriptstyle\!\delta}\textnormal{,max}})\deltat_n),
\end{align}
where $\fc+f_{{\scriptscriptstyle\!\delta}\textnormal{,max}}$ is a heuristic for the worst-case scenario.

\begin{rem}
When the impact of~\fd is ignored, the results in~\fref{eq:4_timeapproxNB} corresponds to the classical approach in the literature to study the effect of delay mismatches on the main beam only~\cite{Vorobyov03,Chen25}.
In such results, \fc is typically considered instead $\fc+f_{{\scriptscriptstyle\!\delta}\textnormal{,max}}$.
\end{rem}

The key takeaway of this section is that the impact of both phase and delay mismatches can be reduced to~\fref{eq:4_replicas} by considering the relative mismatch $\bme=j \bmp$ instead of $\bme=\bmm - \bOne$ as in~\fref{eq:4_mismatchsequence}, where, in the case of a delay mismatch, the phase mismatches in $\bmp$ are approximated as {$p_n\!=\!-2\pi(\fc+f_{{\scriptscriptstyle\!\delta}\textnormal{,max}})\deltat_n$}; this corresponds to a NB approximation.

%% ------------------------------------------------------------------ %%
%% ------------------------------------------------------------------ %%
\subsection{Random Mismatches}
\label{sec:random_mismatches}

We established that the impact of the considered mismatches on the base function is an error term $\Epsilon_\bmm$ consisting of weighted replicas of the original base function.
Those results are useful once a mismatch sequence is known, but mismatches are, in general, unknown to the circuit designer as they are a result of a stochastic process originating during fabrication.
In practice,  designers need to obtain mismatch variances from a system performance target under a production yield requirement: indeed, computing an average of the impact is not informative as one does not want a system to work only on average.
In the following, we will derive a statistical model with the common (among circuit designers) assumption of gain and phase/delay mismatches being independent and following a Gaussian distribution; we will also briefly consider the case of a uniform distribution in \fref{sec:case_of_calibration}.

To gain insight into the statistical behavior of the mismatched beam pattern, we analyze 
\begin{align}
	\Epsilon_\bmm(u) = \sum_{n=0}^{N-1} w_ne_n \mathrm{e}^{-2\pi j un},
\end{align}
the error term in~\fref{eq:4_replicas}. Here, \bme is a random vector containing realizations of i.i.d.\ zero-mean complex Gaussian random variables with independent real and imaginary part of variance~$\sigmar^2$ and~$\sigmai^2$, corresponding to the variance of the gain and phase mismatches, respectively.
In this case, $\Epsilon_\bmm$ is a zero-mean (finite rank) complex Gaussian process~\cite{Neeser93}, fully characterized by its covariance
\begin{align} \label{eq:4_processK}
	K(u, s) = \opE\!\left[\tilde{\Epsilon}(u)\tilde{\Epsilon}(s)^*\right] &= \sum_{n=0}^{N-1} \opE\!\left[\epsilon_n\epsilon_n^*\right] w_n^2 \mathrm{e}^{-2\pi j (u-s)n} \nonumber \\
								&= \left( \sigmar^2 + \sigmai^2\right)Q(u-s)
\end{align}
and pseudo-covariance
\begin{align} \label{eq:4_processJ}
	J(u, s) = \opE\!\left[\tilde{\Epsilon}(u)\tilde{\Epsilon}(s)\right] = \left( \sigmar^2 - \sigmai^2\right)Q(u+s),
\end{align}
where $Q$ is the DTFT of the vector $\tilde{\bmw}$ with its entries squared. 
As the pseudo-covariance $J$ is nonzero and cannot be written as a function of $u-s$, the random process is not proper and not wide-sense stationary, except for the special case~$\sigmar^2=\sigmai^2$.

\begin{rem}
Describing the effect of mismatches on the window~$\tilde{W}$ as a random Gaussian process shows that a statistical framework providing a marginal distribution per angle is of limited use in estimating the impact of mismatches on the SLL. Indeed, the mismatched beampattern shows correlation between different angles, as demonstrated by~\fref{eq:4_processK} and~\fref{eq:4_processJ}. This correlation is intuitive as there are only $2N$ degrees of freedom, corresponding to the real and imaginary part of the mismatch sequence.
 \end{rem}

%% ------------------------------------------------------------------ %%
%% ------------------------------------------------------------------ %%
\begin{figure*}[tp]
	\centering
	\subfigure[SLL CDF with different \sigmar vs. \sigmai distribution]{\includegraphics[width=0.325\linewidth]%
	{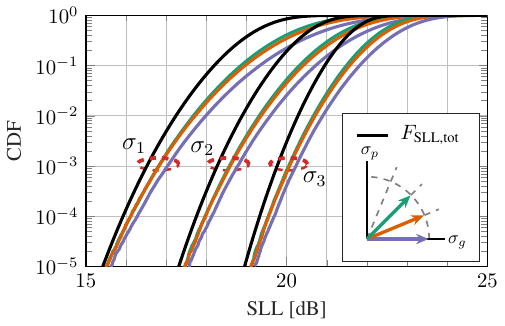}\label{fig:4_cdf_16_angle}}
	\hfill
	\subfigure[SLL CDF  for $N\!\!=\!16$, $25$\,dB-SLL windows]{\includegraphics[width=0.325\linewidth]%
	{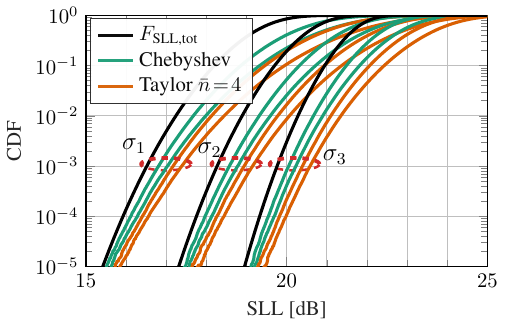}\label{fig:4_cdf_16}}
	\subfigure[SLL CDF  for $N\!\!=\!64$, $30$\,dB-SLL windows]{\includegraphics[width=0.325\linewidth]%
	{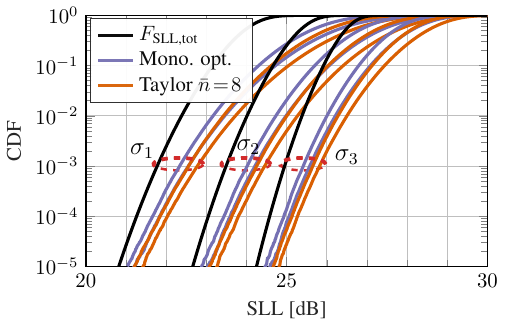}\label{fig:4_cdf_64}}
	\vspace{-0.15cm}  
	\caption{(a) SLL CDF for an $N\!\!=\!16$, $25$\,dB-SLL Chebyshev window for different total mismatch std. deviations $\sigma_1\!=\!0.12$, $\sigma_2\!=\!\sigma_1/\sqrt{2}$, and $\sigma_3\!=\!\sigma_1/2$; the CDF for different distribution between \sigmar and \sigmai shows the relevance of the circularly-symmetric approximation at low quantile, %
	(b) and (c) SLL CDF of windows targeting $25$\,dB-SLL with $N\!\!=\!16$ and $30$\,dB-SLL with $N\!\!=\!64$, showing the quality of the proposed approximation at relevant $\leq\!10^{-3}\!$ quantiles.} 
	\label{fig:4_cdfs}
\end{figure*}

\begin{figure*}[tp]	
	\centering
	\subfigure[Realizations of mismatched windows]{\includegraphics[width=0.325\linewidth]%
	{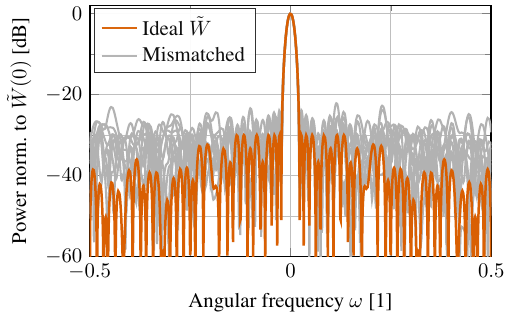}\label{fig:4_window_mis_real}}
	\hfill
	\subfigure[Required $\SLL_{\tilde{W}}$ vs. allowed delay mismatch~$\sigma_\textnormal{t}$]{\includegraphics[width=0.325\linewidth]%
	{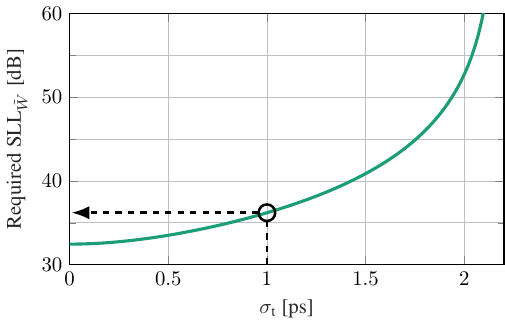}\label{fig:4_target}}
	\hfill
	\subfigure[SLL CDF for the $N\!\!=\!256$ example]{\includegraphics[width=0.325\linewidth]%
	{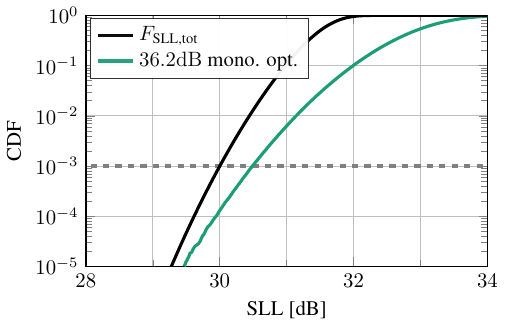}\label{fig:4_cdf_256}}

	\vspace{-0.15cm}  
	\caption{(a) Realizations of $N\!\!=\!64$ $25$\,dB-SLL monotonic optimal windows with Gaussian mismatches $\sigmar\!=\sigmai\!=0.12$, %
	(b) required $N\!\!=\!256$ tapering window SLL as a function of the allowed delay mismatch standard-deviation $\sigma_\textnormal{t}$ based on our approximation with $1$\,dB gain calibration steps, and (c) the obtained monotonic optimal window targeting $30$\,dB-SLL at $q=10^{-3}$ miss rate for the chosen $\sigma_\textnormal{t}=1$\,ps.} 
	\label{fig:4_example}
\end{figure*}

%% ------------------------------------------------------------------ %%
%% ------------------------------------------------------------------ %%
\subsection{Approximation of the SLL Statistics} \label{sec:4_sllstat}

Under the Gaussian assumption, the statistical characterization in \fref{sec:random_mismatches} are exact, but do not translate easily to a distribution of the SLL for a given pair of gain and phase variances $\left(\sigmar^2, \sigmai^2\right)$.
In order to gain more insight, we make two assumptions.
First, we assume that the SLL limit set by mismatches corresponds to the amplitude~$R$ of the strongest replica in~\fref{eq:4_replicas} defined as
\begin{align} \label{eq:4_defR}
	R \triangleq \max_{n \in \llbracket 1,N-1 \rrbracket} |r_n|.
\end{align}
Second, in \fref{eq:4_defR}, we set $\bmr=\bF\bme_\textnormal{tot}$ with the assumption that $\bme_\textnormal{tot}$ contains realizations of i.i.d.\ circularly-symmetric Gaussian random variables of variance $\sigmatot^2 \triangleq \sigmar^2+\sigmai^2$.
We focus on this circularly-symmetric case to obtain tractable expressions, and we will show (in this section) that it serves as a good approximation in the general case by providing results of Monte--Carlo simulations.

\begin{rem}
In \fref{eq:4_defR}, we ignore the impact of the replica of index {$n=0$} which is aligned with the wanted window. Indeed, this replica has a minimal impact on the SLL as it corresponds to scaling the original window by a factor corresponding to a mismatch and hence being negligible compared to $1$; this effect is illustrated in~\fref{fig:4_window_mis_real} where we can see that the MLL is barely affected by mismatches.
\end{rem}

The DFT \bmr of a vector containing i.i.d circularly-symmetric Gaussians, as defined in~\fref{eq:4_defR}, contains realizations of i.i.d circularly-symmetric Gaussians of variance $\sigmatot^2/N$; the magnitude of each entry of $\bmr$ is then Rayleigh distributed.
As those magnitude entries are mutually independent, the CDF of the maximum entry over the vector is the product of the individual entry CDFs
\begin{align} \label{eq:4_Rcdf}
	F_R(z) = \!\left(1-\mathrm{e}^\frac{-Nz^2}{\sigmatot^2}\right)^{\!\!N-1}\!, z\geq0,
\end{align}
where $z$ is the magnitude of the strongest replicas.

We approximate the SLL of the mismatched window function~by  
\begin{align} \label{eq:4_SLLtotdefinition}
	\SLL_\textnormal{tot} \define \frac{1}{1/ \SLL_{\tilde{W}} + R},
\end{align}
 which simply adds up the magnitude $R$ of the strongest replica with the magnitude of the strongest side lobe already present in the wanted window $\tilde{W}$ determined by $\SLL_{\tilde{W}}$. 
With this definition, we obtain the CDF of $\SLL_\textnormal{tot}$ as
\begin{align} \label{eq:4_SLLcdf}
	F_{\SLL_\textnormal{tot}}(z)=1-F_R\!\left( \frac{1}{z} - \frac{1}{\SLL_{\tilde{W}}} \right)\!,
\end{align}
with $0<z<\SLL_{\tilde{W}}$. The upper limit on the values of $z$ translates the fact that we only model the SLL decreasing due to mismatches.
Solving for $z$ in~\fref{eq:4_SLLcdf} with \fref{eq:4_Rcdf} leads to the target SLL not met with a miss rate $q$ given by  
\begin{align} \label{eq:4_SLLq}
	\SLL_{\textnormal{tot,}q} = \!\left( \frac{1}{\SLL_{\tilde{W}}}  +   \sigmatot\alpha_{N,q}    \right)^{-1}.
\end{align}
Here, $\alpha_{N,q}$ is a factor fixed by the number of antenna $N$ and the target miss rate $q$ only
\begin{align} \label{eq:defalpha}
	\alpha_{N,q} \triangleq \sqrt{\frac{-\ln\!\left(  1-(1-q)^{\frac{1}{N-1}}    \right)}{N}}.
\end{align} 
In \fref{eq:defalpha}, $1-q$ corresponds to the production yield. 

\begin{rem}
	The expression in~\fref{eq:defalpha} reveals that, for any relevant values of miss rate $q<10^{-1}$ and number of antennas {$N>2$}, $\alpha_{N,q}$ decreases when the number of antenna $N$ increases at a fixed miss rate target $q$.
	By applying this reasoning in~\fref{eq:4_SLLq}, we conclude that increasing the number of antennas while keeping the same mismatch variance per antenna \sigmatot and the same ideal window $\SLL_{\tilde{W}}$ reduces the impact of mismatches on the SLL of the mismatched window.
\end{rem}

In \fref{fig:4_cdfs}, we demonstrate the relevance of the approximation~$\SLL_\textnormal{tot}$ as defined in \fref{eq:4_SLLtotdefinition}, by comparing its CDF from~\fref{eq:4_SLLcdf} to Monte--Carlo simulations where gain and phase mismatches are mutually independent and contain realizations of i.i.d.\ Gaussian random variables with variances $\sigmar^2$ and $\sigmai^2$, respectively.

\fref{fig:4_cdf_16_angle} shows the CDF of the SLL for an $N=16$ antenna $25$\,dB-SLL Chebyshev window for three different total standard-deviations $\sigmatot$ 
given by $\sigma_1\!=\!0.12$, $\sigma_2\!=\!\sigma_1/\sqrt{2}$, and $\sigma_3\!=\!\sigma_1/2$.
For each of those values for \sigmatot, we distribute the mismatch variance (i) equally, corresponding to $45^\circ$ on the $\sigmar/\sigmai$ plane, (ii) with $\sigmar>0$ but $\sigmai=0$ ($0^\circ$ on the $\sigmar/\sigmai$ plane), and (iii) with the intermediate $22.5^\circ$ case (on the $\sigmar/\sigmai$ plane). We note that, by symmetry, the cases of $67.5^\circ$ and $90^\circ$ are equivalent to $22.5^\circ$ and $0^\circ$, respectively.
Our simulation results show that the variance distribution between \sigmar and \sigmai at a fixed \sigmatot has an influence in the order of $0.5$\,dB already with quantiles as large a $10^{-1}$, and tends to zero when the considered quantile decreases. 

\fref{fig:4_cdf_16} shows the SLL CDF with the same three \sigmatot as above for the $N=16$ antenna $25\,$dB-SLL Chebyshev window and the $\tilde{n}=4$ Taylor window targeting the same SLL.\footnote{For fairness of comparison, the Taylor windows we show correspond to the windows reaching precisely the claimed SLL; the SLL set to generate them is fine-tuned so that this condition is satisfied.}
\fref{fig:4_cdf_64} shows the case of $N=64$ antenna $30$\,dB-SLL $\tilde{n}=8$ Taylor window and its monotonic optimal window counterpart.
In \fref{fig:4_cdf_16} and \fref{fig:4_cdf_64}, we show the two $0^\circ$ and $45^\circ$ extreme distributions of the total variance between $\sigmar$ and $\sigmai$ for each window at each \sigmatot.
The proposed approximation provides an estimate of the CDF for quantiles not exceeding $10^{-3}$ within $1$\,dB of error in all of our simulations, hence enabling quick exploration of the design spaces that the computational complexity of Monte--Carlo methods do not allow.

\begin{rem}
We observe from~\fref{fig:4_cdfs} that all window functions do not perform equally under the same exact mismatch distribution.
This fact is intuitive under our replica interpretation: in regimes where mismatches have little impact on the SLL, a weak replica of the main lobe has to constructively add to an existing secondary lobe to have an impact.
Windows having many secondary lobes close to the power limiting the SLL will be more likely to be impacted.
Indeed, we observe that the equiripple Chebyshev windows and the monotonic optimal windows, which tends to a Chebyshev window with larger target SLL, statistically have worse SLL than their Taylor window counterparts, which have decaying side lobes, under similar mismatch conditions.
However, when the impact of mismatches is stronger, i.e., at lower quantiles in our CDFs, the SLL is limited by one or more replicas that are significantly larger than the window original secondary lobes. Whether a lobe previously limiting the SLL was already present there or not has little impact on the resulting SLL. 
In this last regime, the nature of the original window does not matter much, and we indeed observe in our simulations that all of the considered windows perform similarly at lower quantiles.
\end{rem}

%% ------------------------------------------------------------------ %%
%% ------------------------------------------------------------------ %%
\subsection{Case of Calibration}
\label{sec:case_of_calibration}

To mitigate the impact of mismatches, circuit implementations of amplifiers often have calibration capabilities.
For a relative gain calibration step size of $\Delta_\textnormal{g}$, the resulting relative gain error (after calibration) can be modeled as a uniform distribution of support $\left[-\Delta_\textnormal{g}/2,\Delta_\textnormal{g}/2 \right]$.
As $N$ grows, by the central limit theorem, the distribution of $R$ originating from uniformly distributed mismatches approaches the Gaussian case in~\fref{sec:4_sllstat} with $\sigmatot^2=\Delta_\textnormal{g}^2/12$.
Typically, the Gaussian approximation leads to CDFs with errors of less than $1$\,dB at quantiles as low as $10^{-4}$ for $N=16$~\cite{Guichemerre26}.
The impact of a hypothetical phase/delay calibration scheme can be analyzed analogously. 

%% ------------------------------------------------------------------ %%
%% ------------------------------------------------------------------ %%
\subsection{Design Example} \label{sec:4_exampledesign}

We now illustrate the usefulness of the proposed analysis in obtaining design specifications.
We consider the case of a $256$-antenna phased array transmitting at a carrier frequency of $\fc=10\GHz$ with an RF bandwidth of $\mathit{BW}\!=400\MHz$; we wish to design a tapering window that meets a target of $\SLL_{\textnormal{tot,}q}=30$\,dB with a miss rate of $q=10^{-3}$ (which corresponds to a yield of $1-10^{-3}$).
We assume that gains can be calibrated with $\Delta_\text{dB}=1$\,dB steps, and we want to determine the allowed delay mismatch on the RF signal with variance~$\sigmat^2$.
We still have a degree of freedom, which is the tradeoff between~$\SLL_{\tilde{W}}$, the SLL of the window without mismatch and the delay mismatch~$\sigmat^2$.
With these assumptions, the total mismatch variance is given by
\begin{align} \label{eq:4_sigmatot}
	\sigmatot^2 = \frac{1}{12}\!\left(  10^\frac{\Delta_{\text{dB}}}{20} -1 \right)^2 \!+ 4\pi^2\fmax^2\sigmat^2,
\end{align}
with $\fmax=\fc+\mathit{BW}\!/2$.
Solving for $\SLL_{\tilde{W}}$ in~\fref{eq:4_SLLq} and inserting~\fref{eq:4_sigmatot} leads to the required $\SLL_{\tilde{W}}$ for a given $\sigmat$.

The resulting tradeoff between required $\SLL_{\tilde{W}}$ and mismatch variance \sigmat is shown in~\fref{fig:4_target}. 
When \sigmat is close to zero, the required $\SLL_{\tilde{W}}$ tends to approximatively $32.5\,$dB, which is the limit set by our fixed gain mismatch variance~\sigmar.
When \sigmat is increasing, the required $\SLL_{\tilde{W}}$ tends to infinity when the SLL limit set by \sigmat approaches the target $\SLL_{\textnormal{tot,}q}$.
Furthermore, $\SLL_{\textnormal{tot,}q}$ becomes unachievable when $\sigmatot \geq 1/(\SLL_{\textnormal{tot,}q}\alpha_{N,q})$.

Setting the requirement of a very small \sigmat allows to limit losses in MLL as the required $\SLL_{\tilde{W}}$ is kept low. 
However, designers are limited by the practical achievability set by their choice in \sigmat.
To limit the MLL loss while targeting a practically achievable delay mismatch variance, we choose the pair $\sigmatot=1$\,ps and $\SLL_{\tilde{W}}=36.2\,$dB.
\fref{fig:4_cdf_256} show the result of a Monte--Carlo simulation for uniformly-distributed gain mismatches and Gaussian phase mismatches as derived above. These results reveal that the SLL achieved by a $36.2$\,dB-SLL monotonic optimal window at a miss rate $q=10^{-3}$ is $30.5\,$dB, $0.5$\,dB higher than the $30\,$dB value predicted by our model in \fref{sec:4_sllstat}.

% !TEX root = ../Beam_mismatch.tex
% DO NOT REMOVE THE ABOVE COMMENT!

\section{Extension to UPAs}
\label{sec:UPAextension}

We now demonstrate that the analysis carried out for ULA can be extended to a 2D uniform planar array. 
For our analysis, we need the 2-dimensional ($2$D) DTFT of a two-indices complex sequence $\{u_{n_x,n_y}\}_{n_x,n_y\in\integers}$:
\begin{align}
	U(\omega_x,\omega_y) = \sum_{n_x=-\infty}^{+\infty}\sum_{n_y=-\infty}^{+\infty} u_{n_x,n_y} \mathrm{e}^{-2\pi j (n_x\omega_x+n_y\omega_y)}.
\end{align}
The associated circular convolution is given by
\begin{align} \label{eq:2Dcircconv}
	\left(U\circledast V\right) &(\omega_x,\omega_y) \nonumber \\
	 = & \int_{0}^{1}  \!\!\int_{0}^{1}\!\!U(\tau_x, \tau_y)
	V(\omega_x-\tau_x, \omega_y-\tau_y) \mathrm{d}\tau_x\mathrm{d}\tau_y.
\end{align}

\begin{figure*}[tp]	
	\centering
	\subfigure[]{\includegraphics[width=0.45\linewidth]%
	{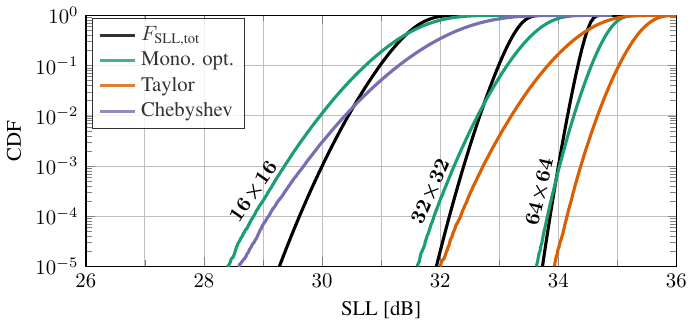}\label{fig:6_2Dcdfmismecc_low}}
	\hspace{1cm}
	\subfigure[]{\includegraphics[width=0.45\linewidth]%
	{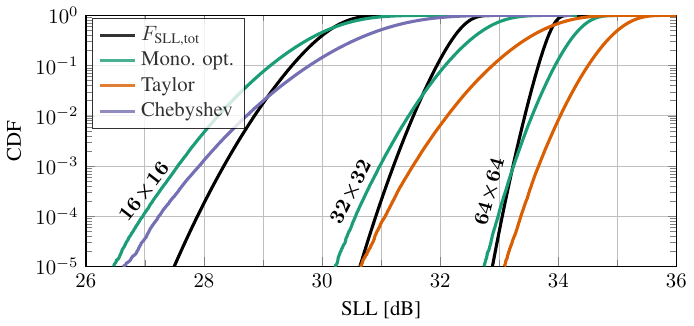}\label{fig:6_2Dcdfmismecc_high}}

	\vspace{-0.15cm}  
	\caption{SLL CDF for different square UPAs and various $36.2$dB-SLL windows (Taylor windows are set to $\bar{n}=8$ and  $\bar{n}=12$ for $32\!\times\!32$ and $64\!\times\!64$ UPA, respectively) with
	(a) uniformly-distributed gain mismatches and Gaussian phase mismatches as derived in~\fref{sec:4_exampledesign}, and (b) double $\sigmar^2$ and $\sigmai^2$ compared to (a).} 
	\label{fig:6_2Dcdfmismecc}
\end{figure*}

%% ------------------------------------------------------------------ %%
%% ------------------------------------------------------------------ %%
\subsection{Framework}

We consider an $N_x\times N_y$ uniform planar array (UPA) with antenna placed along the $x$ and $y$ dimensions, spaced by $\Delta_x'$ and $\Delta_y'$, respectively.
We further define the normalized antenna spacings $\Delta_x=\Delta_x'/\lambda$ and $\Delta_y=\Delta_y'/\lambda$.

Similar to~\fref{sec:beamttdmodel}, the UPA steers the baseband signal $s(t)$ toward the elevation steering angle $\theta_\textnormal{el}\in[0^\circ,90^\circ]$, where $\theta_\textnormal{el}=0^\circ$ corresponds to the array boresight, and the azimuth steering angle $\theta_\textnormal{az}\in]-180^\circ,180^\circ]$, where $\theta_\textnormal{az}=0^\circ$ corresponds to beamforming along the $x$-axis.
We then probe the far-field effectively transmitted signal $r(\varphi_\textnormal{el},\varphi_\textnormal{az}, t)$ at the elevation and azimuth probing angles  $\varphi_\textnormal{el}$ and $\varphi_\textnormal{az}$.

In what follows, we also use the $uv$-coordinate defined as
\begin{align} \label{eq:defuvcoord}
	\begin{cases}
		u_\textnormal{s} = \sin(\theta_\textnormal{el})\cos(\theta_\textnormal{az}) \\
		v_\textnormal{s} = \sin(\theta_\textnormal{el})\sin(\theta_\textnormal{az})
	\end{cases} 
	\!\!\!\textnormal{and }
	\begin{cases}
		u_\textnormal{p} = \sin(\varphi_\textnormal{el})\cos(\varphi_\textnormal{az}) \\
		v_\textnormal{p} = \sin(\varphi_\textnormal{el})\sin(\varphi_\textnormal{az}). 
	\end{cases} 
\end{align}
The $u$ and $v$ coordinate correspond to the length of the unit vector pointing to the considered elevation and azimuth spherical angles when projected onto the $x$ and $y$ axes, respectively.
In~\fref{eq:defuvcoord}, $(u_\textnormal{s}, v_\textnormal{s} )$ corresponds to the steering direction and $(u_\textnormal{p}, v_\textnormal{p} )$ to the probing direction.

We now rewrite the probed far-field signal from~\fref{eq:2_recombin} as
\begin{align}
	r(&u_\textnormal{p}, v_\textnormal{p}, t) \nonumber \\
	&=\frac{g(u_\textnormal{p}, v_\textnormal{p})}{\ell}   
			\sum_{n_x=0}^{N_x-1}\sum_{n_y=0}^{N_y-1}&& x_{n_x,n_y}\!\left( \textstyle t- \frac{n_x\Delta_xu_\textnormal{p} + n_y\Delta_yv_\textnormal{p}}{\fc}     \right) \nonumber \\ 
			& &&\cdot\mathrm{e}^{  -2\pi j  (n_x\Delta_xu_\textnormal{p} + n_y\Delta_yv_\textnormal{p})   },
\end{align}
where $x_{n_x,n_y}(\cdot)$ is the baseband transmit signal of the antenna of index $(n_x,n_y)$.

Analogous to~\fref{eq:2_TTD_2} and~\fref{eq:2_NB}, TTD and NB beamforming are defined by performing MRT with and without time delay.
In both cases, we apply the FT and obtain the beampattern
\begin{align} \label{eq:2Dbeampattern}
	P_{\!\bA}(u_\textnormal{p}, v_\textnormal{p}, \fd) = \!\left|  \tilde{A}\!\left(  \beta^{u_\textnormal{s}}(u_\textnormal{p}, \fd) , \beta^{v_\textnormal{s}}(v_\textnormal{p}, \fd)      \right) \right|^2,
\end{align}
where the base function $\tilde{A}$ is the DTFT of the matrix $\bA\in[0,1]^{N_x\times N_y}$ containing the array per-antenna gains extended by zero padding.
In the context of TTD beamforming, the deformation function is defined as
\begin{equation}
\begin{aligned}
	\beta_\textnormal{\tiny TTD}^{u_\textnormal{s}}(u_\textnormal{p}, \fd) &= \Delta_x (u_\textnormal{p}-u_\textnormal{s})\!\left(1+ \frac{\fd}{\fc} \right), \\
	\beta_\textnormal{\tiny TTD}^{v_\textnormal{s}}(v_\textnormal{p}, \fd) &= \Delta_y (v_\textnormal{p}-v_\textnormal{s})\!\left(1+ \frac{\fd}{\fc} \right),
\end{aligned}
\end{equation}
and in the context of NB beamforming as
\begin{equation} \label{eq:2DNBbetas}
\begin{aligned}
	\beta_\textnormal{\tiny NB}^{u_\textnormal{s}}(u_\textnormal{p}, \fd) &= \Delta_x \!\left(\!\left(1+ \frac{\fd}{\fc} \right)\!u_\textnormal{p}-u_\textnormal{s}\right), \\
	\beta_\textnormal{\tiny NB}^{v_\textnormal{s}}(v_\textnormal{p}, \fd) &= \Delta_y \!\left(\!\left(1+ \frac{\fd}{\fc} \right)\!v_\textnormal{p}-v_\textnormal{s}\right).
\end{aligned}
\end{equation}
Those results show that, similar to a ULA, a frequency domain representation of a UPA allows to decompose the impact of beamforming into a base and a deformation function.
As visible in~\fref{eq:2DNBbetas}, beamsquint also affects UPAs if NB beamforming is employed.
In this case, solving for zero in~\fref{eq:2DNBbetas} and translating the resulting $uv$-coordinates back to spherical angles shows that the main lobe peak appears at the expected azimuth angle $\theta_\textnormal{az}$, but at frequency-dependent elevation angle
\begin{align}
	\varphi_\textnormal{el,main} = \arcsin\!\left(\frac{\fc}{\fd + \fc} \sin(\theta_\textnormal{el})\! \right).
\end{align}

%% ------------------------------------------------------------------ %%
%% ------------------------------------------------------------------ %%
\subsection{Separable Tapering}

With UPAs, tapering is also required in order to obtain an SLL higher than $13\,$dB.
A classic approach consists in implementing \emph{separable} 2D tapering windows, which are defined as the element-wise product of two 1-dimensional windows, one for the $x$ axis and one for the $y$ axis~ \cite{Mailloux17}.
In this case, the coefficients of the separable tapering window $\bW\!\in[0,1]^{N_x\times N_y}$ can be written as
\begin{align}
	\{ \bW\}_{n_x,n_y} = w^x_{n_x}w^y_{n_y},
\end{align}
where $\bmw^x\!\in[0,1]^{N_x}$ and $\bmw^y\!\in[0,1]^{N_y}$ are $1$D tapering windows associated to the $x$ and $y$ axes, respectively, which leads to the 2D DTFT
\begin{align}
	\tilde{W}(\omega_x,\omega_y) = \tilde{W}^x(\omega_x)\tilde{W}^y(\omega_y),
\end{align}
with $\tilde{W}^x$ and $\tilde{W}^y$ the DTFTs of the zero-padded extensions of $\bmw^x$ and $\bmw^y$, respectively.

%% ------------------------------------------------------------------ %%
%% ------------------------------------------------------------------ %%
\subsection{Non-Separable Tapering}

The separable $2$D windows have the advantage of directly extending their $1$D counterpart, but do not exhibit any kind of MLL vs. SLL optimality. Indeed, the strongest sidelobes of $\tilde{W}$ can only exist when either $\omega_x$ or $\omega_y$ is zero;\footnote{The strongest sidelobe of $\tilde{W}^x$ can only limit $\tilde{W}$ SLL when $\tilde{W}^y$ is at its peak, i.e., when $\omega_y=0$. The same reasoning applies for the strongest sidelobe of $\tilde{W}^y$.} enforcing lower sidelobes than needed on most of the directions leads to a loss in MLL.
The convex optimization formulation in~\fref{sec:tradeoffMLLSLL} can be adapted for the UPA case and yields MLL vs. SLL tradeoff optimal 2D tapering windows. For example, a $64\times64$ separable $30$\,dB-SLL Taylor window with $\bar{n}=8$ yields an MLL of $-7.4\,$dB, whereas the corresponding monotonic optimal\footnote{In this context, a monotonic optimal $2$D window is a window for which each row and column is a $1$D monotonic optimal window.} window achieves $-5.5\,$dB MLL, which corresponds to a $55\%$ increase in main lobe peak power.

%% ------------------------------------------------------------------ %%
%% ------------------------------------------------------------------ %%
\subsection{Mismatch}

In the presence of gain, phase, and delay mismatches, the base function $\tilde{A}$ in~\fref{eq:2Dbeampattern} can be modeled as 
\begin{align}
	\tilde{A}_\bmm = \tilde{W} + \Epsilon_\bmm,
\end{align}
with $\tilde{W}$ the intended $2$D tapering window and the error term
\begin{align}
	\!\!\!\Epsilon_\bmm(\omega_x,\omega_y) = \!\!\sum_{n_x=0}^{N_x-1}\, \sum_{n_y=0}^{N_y-1} \!
	r_{n_x,n_y} \!\tilde{W}\!\left( \omega_x \!-\! \frac{n_x}{N_x}, \omega_y \!-\! \frac{n_y}{N_y}  \right)\!.
\end{align}
Analogous to the ULA case in~\fref{eq:4_replicas}, $\Epsilon_\bmm$ consists of replicas of the intended window weighted by $\{r_{n_x,n_y}\}$, the $2$D DTFT of the (complex) per antenna mismatches.
As the $2$D DFT of i.i.d. $2$D circularly-symmetric Gaussian random variables contains i.i.d. circularly-symmetric Gaussians, the analysis in~\fref{sec:4_sllstat} can be used to approximate the CDF of the SLL in the presence of mismatches by setting $N=N_xN_y$.
The relevance of this approach is shown in~\fref{fig:6_2Dcdfmismecc_low} where the per-antenna mismatch is the same as in the design example in~\fref{sec:4_exampledesign}, and in~\fref{fig:6_2Dcdfmismecc_high} where the variance of the real and imaginary part of the per-antenna mismatches are doubled.
In both cases, the proposed approximation follows the tail behavior of the CDF, which is relevant for yield analysis, with errors lower than $1\,$dB in the relevant $10^{-3}$ to $10^{-5}$ range, and correctly predicts the impact of varying the mismatch variance or the number of antenna.

\begin{rem}
In the 1-dimensional case, the strongest-replica approximation tends to be conservative, as the replica spectra are arranged along a line; the mismatched beampattern at a given location is typically influenced by only one or two nearby shifted copies of the intended tapering window.
In the 2-dimensional case, however, the replicas form a lattice in the spatial-frequency plane, so each point may receive non-negligible contributions from several nearby replicas simultaneously.
As a result, the SLL is governed by the joint interaction of multiple replicas rather than by a single dominant one, which explains why the strongest-replica approximation does not retain its empirically conservative behavior for UPAs.
\end{rem}

% !TEX root = ../Beam_mismatch.tex
% DO NOT REMOVE THE ABOVE COMMENT!

%% ------------------------------------------------------------------ %%
%% ------------------------------------------------------------------ %%
\section{Limitations and Future Work}
\label{sec:limitations}

The framework proposed in~\fref{sec:Framework} and its application on analyzing tapering (\fref{sec:beamtapering}) and the impact of mismatches (\fref{sec:beammismatchsec}) do not include the unavoidable impact of antenna coupling and matching.
Such effects could be studied by applying the proposed framework in conjunction with a physically-consistent antenna-modeling framework, e.g., the one in~\cite{stutz25}.
We also do not study the effect of mismatches on beamfocusing in the near-field, which is definitely an interesting topic for future research.
Furthermore, we only partially address antenna directivity as this effect has no influence on the global implications of our results.

%% ------------------------------------------------------------------ %%
%% ------------------------------------------------------------------ %%
\section{Conclusions}
\label{sec:conclusions}

We have developed a frequency-domain framework for phased-array beamforming that separates the frequency-dependent effects from the array characteristics.
Building upon this independency, we have reintroduced tapering in the lens of the proposed framework.
We have then described the impact of both deterministic and random gain and delay/phase mismatches, and we have proposed an approximation of the impact of random mismatches on the SLL of a given window; the analytical expressions resulting from our approximation serve as a worst-case analysis and enable quick exploration of the design space without the need to perform computationally-intensive  Monte--Carlo simulations.
Using a 256-antenna $10$\,GHz design example, we have demonstrated how the proposed framework translates a target SLL and yield requirement into concrete window and hardware specifications, with Monte--Carlo simulations confirming agreement to within $0.5$\,dB of our analytical prediction.

\balance

\bibliographystyle{IEEEtran}
\bibliography{}

\balance

\end{document}